\documentclass[aps,prd,superscriptaddress,nofootinbib,amsmath,amssymb,showpacs]{revtex4-2}
\usepackage{bm,amssymb,amsfonts,braket,color}
\newcommand{\beqn}{\begin{eqnarray}}
\newcommand{\eeqn}{\end{eqnarray}}

\newcommand{\cor}[1]{{#1}}
\begin{document}

\title{Absence of equilibrium chiral magnetic effect}

\author{M.A. Zubkov\footnote{on leave of absence from Moscow Institute of Physics and Technology, 9, Institutskii per., Dolgoprudny, Moscow Region, 141700, Russia}}
\email{zubkov@itep.ru}
\affiliation{LE STUDIUM, Loire Valley Institute for Advanced Studies,
Tours and Orleans, 45000 Orleans France}
\affiliation{Laboratoire de Math´ematiques et de Physique
Th´eorique, Universit´e de Tours, 37200 Tours, France}
\affiliation{Institute for Theoretical and Experimental Physics, B. Cheremushkinskaya 25, Moscow, 117259, Russia}
\affiliation{Far Eastern Federal University,  School of Biomedicine, 690950 Vladivostok, Russia}
%\affiliation{National Research Nuclear University MEPhI (Moscow Engineering
%Physics Institute), Kashirskoe highway 31, 115409 Moscow, Russia}

\begin{abstract}
We analyse the $3+1$ D equilibrium chiral magnetic effect (CME). We apply derivative expansion to the Wigner transform of the two - point Green function. This technique allows us to express the response of electric current to external electromagnetic field strength through the momentum space topological invariant. We consider  the wide class of the lattice regularizations of quantum field theory (that includes, in particular, the regularization with Wilson fermions) and also certain lattice models of solid state physics (including those of Dirac semimetals). It appears, that in these models the mentioned topological invariant vanishes identically at nonzero chiral chemical potential. That means, that the bulk equilibrium CME is absent in those systems.
\end{abstract}


\date{\today}

\maketitle

% Italic ``theorems''
%\theoremstyle{plain}
\newtheorem{theorem}{Theorem}[section]
\newtheorem{hypothesis}[theorem]{Hypothesis}
\newtheorem{lemma}{Lemma}[section]
\newtheorem{corollary}[lemma]{Corollary}
\newtheorem{proposition}[lemma]{Proposition}
\newtheorem{claim}[lemma]{Claim}

% Roman ``theorems''
%\theoremstyle{definition}
\newtheorem{definition}[lemma]{Definition}
\newtheorem{assumption}{Assumption}

% Humble things: remarks and examples.
%\theoremstyle{remark}
\newtheorem{remark}[lemma]{Remark}
\newtheorem{example}[lemma]{Example}
\newtheorem{problem}[lemma]{Problem}
\newtheorem{exercise}[lemma]{Exercise}

%\newenvironment{nam}[args]{begdef}{enddef}

%\maketitle

\newcommand{\br}{{\bf r}}
\newcommand{\bu}{{\bf \delta}}
\newcommand{\bk}{{\bf k}}
\newcommand{\bq}{{\bf q}}
\def\({\left(}
\def\){\right)}
\def\[{\left[}
\def\]{\right]}

\section{Introduction.}

The chiral magnetic effect (CME) has been widely discussed recently in different contexts both within the continuous quantum field theory and in the condensed matter physics. The CME for the case, when the left - handed and the right - handed fermions are truly separated was first discussed in \cite{Vilenkin}. In the context of quantum field theory the existence of chiral magnetic effect was considered in \cite{CME}, followed by a number of papers (see, for example, \cite{Kharzeev:2013ffa,Kharzeev:2009pj} and references therein). In particular, CME has been discussed using a different technique in the Fermi liquids \cite{SonYamamoto2012}.

The possible existence of the  chiral magnetic contribution to conductivity was proposed in \cite{Nielsen:1983rb}, and was discussed later in a number of papers. The experimental observation of this contribution to conductivity in the recently discovered Dirac semimetals was reported in \cite{ZrTe5}. Notice, that from our point of view such a  chiral magnetic contribution to conductivity should be distinguished from the equilibrium CME \cite{CME}. During the calculation of the chiral magnetic contribution to ordinary conductivity in \cite{Nielsen:1983rb,ZrTe5} the chiral imbalance appears as a pure kinetic phenomenon, and the final expression for the CME current is proportional to the squared magnetic field and, in addition, to electric field. At the same time in the equilibrium CME  the nondissipative current is linear in magnetic field and is predicted to appear without any external electric field. Therefore, the linear response theory to be considered in the present paper, strictly speaking, does not describe the chiral magnetic contribution to conductivity. Thus we will concentrate on the equilibrium CME.

The family of the non - dissipative transport effects being the cousins of the CME has also been widely discussed recently both in the context of the high energy physics and in the context of condensed matter theory \cite{Landsteiner:2012kd,semimetal_effects7,Gorbar:2015wya,Miransky:2015ava,Valgushev:2015pjn,Buividovich:2015ara,Buividovich:2015ara,Buividovich:2014dha,Buividovich:2013hza}. The possible appearance of such effects in the recently discovered Dirac and Weyl semimetals has been considered \cite{semimetal_effects6,semimetal_effects10,semimetal_effects11,semimetal_effects12,semimetal_effects13,Zyuzin:2012tv,tewary}.
In the context of the high energy physics the possibility to observe CME in relativistic heavy - ion collisions was widely discussed (see, for example, \cite{Kharzeev:2015znc,Kharzeev:2009mf,Kharzeev:2013ffa} and references therein). Certain lattice calculations seem to confirm indirectly this possibility \cite{Polikarp}.

In several publications the existence of equilibrium CME was questioned. In particular, in \cite{Valgushev:2015pjn,Buividovich:2015ara,Buividovich:2015ara,Buividovich:2014dha,Buividovich:2013hza} using different numerical methods the CME current was investigated in the context of lattice field theory. It was argued, that the equilibrium bulk CME does not exist, but close to the boundary of the system the nonzero CME current may appear. It was demonstrated, that in the given systems the integrated total CME current remains zero. The similar conclusion was drawn in \cite{Gorbar:2015wya} basing on the consideration of the system of finite size with the special boundary conditions in the direction of the external magnetic field. The consideration of \cite{Gorbar:2015wya}, however, does not refer to the systems, which do not have boundaries or, say, have the form of a circle with magnetic field directed along the circle. In the context of condensed matter theory the absence of CME was reported within the particular model of Weyl semimetal \cite{nogo}. Besides, it was argued, that the equilibrium CME may contradict to the no - go Bloch theorem \cite{nogo2}.

In the present paper we consider CME on the basis of Wigner transformation technique \cite{Wigner,star} applied to  Green functions. First of all, we demonstrate, that the derivative expansion within this technique allows to reduce the expression for the linear response of electric current to the external field strength to the momentum space topological invariant. The power of this method is demonstrated on the example of the 2+1 D quantum Hall effect (QHE), where it allows to derive in a simple way the conventional expression for Hall conductivity \cite{Volovik2003}. Momentum space topology is a powerful method, which was developed earlier mainly within condensed matter theory. In addition to the ordinary quantum Hall effect it allows to describe  in a simple way a lot of the other effects (for the review see \cite{Volovik2003,Volovik:2011kg}). Recently certain aspects of momentum space topology were discussed in the framework of the four - dimensional lattice gauge theory \cite{VZ2012,Z2012}. Here we derive the expression for the linear response of the electric current to the external magnetic field in the wide class of the $3+1$ D fermionic systems, which includes popular lattice regularizations of continuum quantum field theory and the models of discovered recently Dirac semimetals.

Strictly speaking, our calculations remain unambiguous only for the systems with the Green functions that do not have poles (or zeros). It appears, that like in the $2+1$ D case the resulting $3+1$ D response of electric current to the external magnetic field is proportional to the topological invariant in momentum space. Unlike the case of the naive continuum fermions, for the wide class of the lattice regularizations of quantum field theory (and for the certain models of  Dirac semimetals) the mentioned topological invariant vanishes for the nonzero chiral chemical potential. This means, that the equilibrium bulk CME current is absent in the properly regularized quantum field theory and in the discussed models of Dirac semimetals.

The paper is organized as follows. In Sect. \ref{SectCont} we start the discussion of the linear response of electric current to external electromagnetic field using continuum formulation. The Wigner transform of the two - point Green function is defined in Sect. \ref{SectWignCont}. The main equation obeyed by this object is proved in Appendix A. In Section \ref{SectGradCont} we present the gradient expansion for the Wigner transform of the Green function. The linear response of the electric current to external gauge field is considered in Sect. \ref{SectLinCont}. It appears, that the resulting expression is divergent and requires regularization.
In Sect. \ref{SectLat} we consider lattice regularization. In Sect. \ref{SectGaugeLat} we discuss lattice theory of general form, which allows to describe not only the lattice regularization of the continuum quantum field theory, but also the tight - binding like models of solid state physics. We propose the unusual way to introduce the external gauge field to the lattice model. This method allows to deal with the theory written in momentum space, which is important for our further considerations. The proposed method is manifestly gauge invariant, and it is obviously reduced to the conventional minimal connection of theory with the gauge field in continuum limit. Therefore, it allows to introduce effectively the gauge field both into the lattice regularization of quantum field theory and to the models of the solid state physics. In Sect. \ref{SectWignLat} the Wigner transform of the lattice two - point Green function in momentum space is discussed. It appears, that it obeys the same equation as its continuum counterpart. This is proved in Appendix B.  In Sect. \ref{SectLinLat} the linear response of the electric current to the external gauge field is derived for the lattice theory. It appears, that the resulting expression represents the direct lattice discretization of the corresponding continuum expression as expected. This expression is, in turn, a topological invariant in momentum space. This is proved in Appendix C. In Sect. \ref{SectHall} the celebrated expression for the Hall current is reproduced using the proposed technique. In Sect. \ref{SectCME} we finally discuss the chiral magnetic effect.  In Sect. \ref{Sectmu5} we consider the introduction of the chiral chemical potential into the Green function. In Sect. \ref{Sectmu51} the conventional massive lattice fermions are considered while in Sect. \ref{Sectmu52} the marginal models of massive lattice fermions are discussed. In Sect. \ref{Sectmassless} the models with massless fermions are considered. We demonstrated, that in all considered models the bulk CME is absent.  In Sect. \ref{SectConcl} we end with the conclusions.

\section{Continuum theory}
\label{SectCont}

\subsection{Wigner transform of the Green function.}
\label{SectWignCont}

In the present section we recall some of the basic notions of the Wigner (Weyl) transform.  For the more details see, for example, \cite{berezin} and Appendix B in  \cite{Weyl}. Next, we will apply those notions to the two point Green function of a non - interacting fermion system in the presence of external gauge field.

Let us consider the $d+1 = D$ dimensional continuum model with the fermionic Green function ${\cal G}({\bf p})$ that depends on the $D$ vector ${\bf p} = (p_1,...,p_D)$ of Euclidean momentum. (The Wick rotation has been performed.) When interactions between the fermions are neglected, the external electromagnetic field ${\bf A}({\bf r})$ may be taken into account through the Hermitian operator - valued function $\hat{\cal Q}({\bf r},\hat{\bf p}) = {\cal G}^{-1}(\hat{\bf p}-{\bf A}({\bf r}))$, where $\hat{\bf p}= - i \partial_{\bf r}$. Operators $\hat{p}_i - A_i({\bf r})$ and $\hat{p}_j - A_j({\bf r})$ do not commute for $i \ne j$. Therefore, we should point out the way of their ordering inside $\hat{\cal Q}$. We choose the following way for definiteness:  each product $p_{i_1} ... p_{i_n}$ in the expansion of ${\cal G}^{-1}$ is substituted by the symmetrized product $\frac{1}{n!}\sum_{\rm permutations} (\hat{p}_{i_1}-A_{i_1}) ... (\hat{p}_{i_n}-A_{i_n})$. This way of ordering corresponds to the so - called symmetrical (or, Wigner) quantization according to \cite{berezin}.   The resulting function $\hat{\cal Q}$ enters the functional integral representation for the Euclidean partition function
\begin{equation}
Z = \int D\bar{\Psi}D\Psi \, {\rm exp}\Big( - \int d^D {\bf r} \bar{\Psi}({\bf r})\hat{Q}({\bf r},\hat{\bf p})\Psi({\bf r}) \Big)
\end{equation}
Here $\Psi$, $\bar{\Psi}$ are the Grassmann - valued continuum fermionic fields. In the presence of the gauge field the Green function appears as a correlator
\begin{eqnarray}
G({\bf r}_1,{\bf r}_2)&=& \cor{-}\frac{1}{Z}\int D\bar{\Psi}D\Psi \,\bar{\Psi}({\bf r}_2)\Psi({\bf r}_1)\nonumber\\ &&{\rm exp}\Big( - \int d^D {\bf r} \bar{\Psi}({\bf r})\hat{Q}({\bf r},\hat{\bf p})\Psi({\bf r}) \Big)
\end{eqnarray}
It  obeys equation
\begin{equation}
\hat{\cal Q}({\bf r}_1,-i \partial_{{\bf r}_1})G({\bf r}_1,{\bf r}_2) =  \delta^{(D)}({\bf r}_1-{\bf r}_2)\label{QG}
\end{equation}
Wigner transform \cite{Wigner} of the Green function is defined as
\begin{equation}
 \tilde{G}({\bf R},{\bf p}) = \int d^D{\bf r} e^{-i {\bf p} {\bf r}} G({\bf R}+{\bf r}/2,{\bf R}-{\bf r}/2)\label{W}
\end{equation}
In Appendix A the Groenewold equation \cite{Weyl} for the function $\tilde{G}$ is derived
\begin{eqnarray}
1 &=& {\cal Q}({\bf R},{\bf p})*\tilde G({\bf R},{\bf p})\nonumber\\ && \equiv {\cal Q}({\bf R},{\bf p})e^{\frac{i}{2}(\overleftarrow{\partial}_{\bf R}\overrightarrow{\partial}_{\bf p} - \overleftarrow{\partial}_{\bf p}\overrightarrow{\partial}_{\bf R})}\tilde G({\bf R},{\bf p})  \label{id000}
\end{eqnarray}
Here function $\cal Q$ represents the so - called Weyl symbol of operator $\hat{\cal Q}$ being the Wigner transform of its matrix elements \cite{Weyl,berezin}. It depends on the real numbers rather than on the operators. The  explicit form of the relation between $\cal Q$ and $\hat{\cal Q}$ is given in Appendix A. Here we will need only the following property of the correspondence between $\hat{\cal Q}$ and $\cal Q$. If $\hat{\cal Q}$ has the form of a function ${\cal G}^{-1}$ of the combination $(\hat{\bf p} - {\bf A}({\bf r}))$ with a gauge potential ${\bf A}({\bf r})$, i.e.
\begin{equation}
\hat{\cal Q}({\bf r},\hat{\bf p}) = {\cal G}^{-1}(\hat{\bf p} - {\bf A}({\bf r}))\label{Q000}
\end{equation}
then we have
\begin{equation}
{\cal Q}({\bf r},{\bf p}) = {\cal G}^{-1}({\bf p} - {\bf A}({\bf r})) + O([\partial_i A_j]^2)\label{QF}
\end{equation}
Here $O([\partial_i A_j]^2)$ may contain the terms with the second power of the derivative of $\bf A$, with the squared derivative of $\bf A$, and the terms higher order in derivatives.
In principle, the restrictions on the term $O([\partial_i A_j]^2)$ may be more strong\footnote{For example, the Weyl symbol of the operator $f(-i\partial_{r} - Hr)$ for the one - dimensional problem ($r \in {\bf R}^1$) is given by $f(p-Hr)$ exactly \cite{berezin}.}, but for our purposes it will be enough, that the terms linear in the derivatives of $\bf A$ are absent in  $O([\partial_i A_j]^2)$ (which is proved in Appendix A).

Notice, that the star product entering Eq. (\ref{id000}) is widely used in deformation quantisation \cite{berezin,star2} and also in some other applications (see, for example, \cite{star} and references therein). It is also worth mentioning, that the Wigner transformation of the Green function was used in a number of applications. In particular, it was applied to the derivation of quantum kinetic equations \cite{ke,ke2}. The methods to solve kinetic equations that operate with the Wigner transform of the Green function were discussed in \cite{ke3}.

\subsection{Gradient expansion for the Wigner transform of the Green function in the presence of external gauge field.}
\label{SectGradCont}

Here we apply the formalism of Wigner transform to the $d+1= $  D dimensional fermionic systems.
Let us consider the model with the Green function ${\cal G}({\bf p})$ that depends on the D - vector ${\bf p} = (p_1,p_2,...,p_4)$ of Euclidean momentum. We introduce the slowly varying external $U(1)$ vector gauge field ${\bf A}({\bf r})$ defining operator function $\hat{\cal Q}$ of Eq. (\ref{Q000}).
The Wigner transform of the Green function Eq. (\ref{W}) satisfies Eq. (\ref{id000}).

We apply the gradient expansion and come to
\begin{eqnarray}
\tilde G({\bf R},{\bf p})  &= &\tilde G^{(0)}({\bf R},{\bf p}) + \tilde G^{(1)}({\bf R},{\bf p}) + ...  \label{Gexp}\\
\tilde G^{(1)}  &= &\cor{+}\frac{i}{2} \tilde G^{(0)} \frac{\partial \Big[\tilde G^{(0)}\Big]^{-1}}{\partial p_i} \tilde G^{(0)}  \frac{\partial  \Big[\tilde G^{(0)}\Big]^{-1}}{\partial p_j} \tilde G^{(0)}
A_{ij} ({\bf R})\nonumber
\end{eqnarray}
Here $\tilde G^{(0)}({\bf R},{\bf p})$ is defined as the Green function with the field strength $A_{ij} = \partial_i A_j - \partial_j A_i$ neglected. It is given by
\begin{eqnarray}
&&\tilde G^{(0)}({\bf R},{\bf p})  = {\cal G}({\bf p}-{\bf A}({\bf R}))\label{Q0}
\end{eqnarray}

\subsection{Linear response of electric current to the strength of external gauge field.}
\label{SectLinCont}

The components of vector $U(1)$ current in the system of non - interacting fermions may be expressed as:
\begin{eqnarray}
j^{k}({\bf R}) &=& \cor{+}{\rm Tr}\, \frac{\partial  {\cal G}^{-1}(-i\partial_{{\bf r}_1}-{\bf A}({\bf r}_1))}{\partial A_k}\, G( {\bf r}_1, {\bf r}_2)\Big|_{{\bf r}_1,{\bf r}_2\rightarrow {\bf R}}\nonumber\\ &=& \cor{-}\int \frac{d^D {\bf p}}{(2\pi)^D}\,  {\rm Tr}\, \tilde G({\bf R},{\bf p})\frac{\partial  \Big[\tilde G^{(0)}({\bf R},{\bf p})\Big]^{-1}}{\partial p_k}\label{j14}
\end{eqnarray}
For the derivation of the second row in this expression we applied expressions of Appendix A. Also, this expression follows as a continuum limit of the corresponding formula to be derived in the next section. Besides, we advise the reader to consult Appendix B of \cite{Weyl}, where many useful relations are collected, including those, which give rise to Eq. (\ref{j14}).

In the $3+1$ D systems the contribution to electric current originated from $\tilde{G}^{(1)}$ is given by
\begin{eqnarray}
j^{(1)k}({\bf R})  &= &\cor{-} \frac{1}{4\pi^2}\epsilon^{ijkl} {\cal M}_{l} A_{ij} ({\bf R}), \label{calM0}\\
{\cal M}_l &=& \int_{} \,{\rm Tr}\, \nu_{l} \,d^4p \label{Ml} \label{nuG0} \\ \nu_{l} & = &  - \frac{i}{3!\,8\pi^2}\,\epsilon_{ijkl}\, \Big[  {\cal G} \frac{\partial {\cal G}^{-1}}{\partial p_i} \frac{\partial  {\cal G}}{\partial p_j} \frac{\partial  {\cal G}^{-1}}{\partial p_k} \Big]  \label{nuG02}
\end{eqnarray}
In the linear response theory we should substitute ${\bf A}=0$ into the expression for ${\cal M}_l$. Therefore, in Eq. (\ref{nuG0}) we substitute $\cal G$ instead of $\tilde{G}^{(0)}$.

Notice, that our conventions of notations assume, that the field strength absorbs the elementary charge $e$, i.e. the Green function $\cal G$ contains the combination ${\bf p} -{\bf A}$ instead of the conventional ${\bf p}-e {\bf A}$. Besides, the electric current $j$ is defined in the units of $e$, which eliminates the second factor $e$ from Eq. (\ref{calM0}).

In order to understand how Eq. (\ref{nuG0}) works let us consider the single massless Dirac fermion, which is the couple of the left - handed and the right - handed Weyl fermions. The corresponding expression for the Green function in the presence of chiral chemical potential $\mu_5$ is given by
 \begin{equation}
 {\cal G}^{}({\bf p}) = \Big(\sum_{k}\gamma^{k} p_{k} + i\gamma^4 \gamma^5 \mu_5\Big)^{-1}\label{G2W}
 \end{equation}
with the Euclidean gamma - matrices that satisfy $\{\gamma^a,\gamma^b\}=2\delta^{ab}$ and the $\gamma^5$ matrix given by ${\rm diag}(1,1,-1,-1)$ in chiral representation.
In this situation the Green function contains poles. Besides, the integral in Eq. (\ref{nuG0}) is divergent at infinite values of $\bf p$. Therefore, in order to apply the above expressions the regularization is needed.

The obvious expectation about Eq. (\ref{nuG02}) is that in lattice regularization we need simply to substitute into it the lattice Green function defined as a function of lattice momentum, and integrate in Eq. (\ref{nuG0}) over the compact momentum space $\cal M$. Below we will see, that this is indeed what should be done.

\section{Lattice regularized theory}
\label{SectLat}

\subsection{A way to introduce the external gauge field to lattice model.}
\label{SectGaugeLat}

Again, let us consider the $d+1 = D$ dimensional model with the fermionic Green function ${\cal G}({\bf p})$ that depends on the $D$ vector ${\bf p} = (p_1,...,p_D)$ of Euclidean momentum. Now we assume, that momentum space is compact and has the form of the product ${\cal M} = S^1 \otimes \Omega$, where $\Omega$ is the $d$ - dimensional Brillouin zone while $S^1$ is the circle of $p^D$. Momentum space of such form is typical for the lattice regularization of quantum field theory. It also appears in the tight - binding models of condensed matter systems when evolution in time is discretized (which is always necessary to make Monte - Carlo simulations of such systems).  Notice, that the lattice momentum ${\bf p}$ does not appear as the eigenvalue of the operator $-i\partial_{\bf r}$. The same refers also to the solid state models.

In the absence of the external electromagnetic field  the partition function of the theory under consideration may be written as
\begin{equation}
Z = \int D\bar{\Psi}D\Psi \, {\rm exp}\Big( - \int_{\cal M} \frac{d^D {\bf p}}{|{\cal M}|} \bar{\Psi}({\bf p}){\cal G}^{-1}({\bf p})\Psi({\bf p}) \Big)\label{Z1}
\end{equation}
where $|{\cal M}|$ is the volume of momentum space $\cal M$. (We neglect here those interactions, which are not taken into account by the form of the two point Green function ${\cal G}({\bf p})$. Otherwise, the interaction terms with higher powers of $\Psi$ or with the additional dynamical fields would have been written.)

We assume, that the theory to be dealt with has the form of the lattice regularization of the continuum quantum field theory, or the form of the solid state tight - binding like model. In both cases the theory is defined in discrete coordinate space. We assume, that the dynamical variables $\Psi$ of this theory are attached to the lattice sites ${\bf r}_n$.

The fields in coordinate space are related to the fields in momentum space as follows
\begin{equation}
\Psi({\bf r}) = \int_{\cal M} \frac{d^D {\bf p}}{|{\cal M}|} e^{i {\bf p}{\bf r}} \Psi({\bf p})\label{Psip}
\end{equation}
At the discrete values of $\bf r$ corresponding to the points of the lattice this expression gives the values of the fermionic field at these points, i.e. the dynamical variables of the original lattice model. However, Eq. (\ref{Psip}) allows to define formally the values of fields at any other values of $\bf r$. The partition function may be rewritten in the form
\begin{equation}
Z = \int D\bar{\Psi}D\Psi \, {\rm exp}\Big( - \sum_{{\bf r}_n} \bar{\Psi}({\bf r}_n)\Big[{\cal G}^{-1}(-i\partial_{\bf r})\Psi({\bf r})\Big]_{{\bf r}={\bf r}_n} \Big)\label{Z2}
\end{equation}
Here the sum in the exponent is over the discrete coordinates ${\bf r}_n$. However, the operator $-i\partial_{\bf r}$ acts on the function $\Psi({\bf r})$ defined using Eq. (\ref{Psip}). In order to derive Eq. (\ref{Z2}) we use identity
\begin{equation}
\sum_{\bf r}e^{i{\bf p}{\bf r}} = |{\cal M}|\delta({\bf p})
\end{equation}
Gauge transformation of the original lattice field
\begin{equation}
\Psi({\bf r}_n)\rightarrow e^{i \alpha({\bf r}_n)} \Psi({\bf r}_n)
\end{equation}
now may be understood as the gauge transformation of the field $\Psi$ defined for any values of $\bf r$: we simply extend the definition of the function $\alpha({\bf r})$ to the function, which is defined at any values of $\bf r$ and take the original values at the discrete lattice points. This prompts the following way to introduce the external gauge field to our lattice model. We consider the partition function of the form
\begin{eqnarray}
Z &=& \int D\bar{\Psi}D\Psi \, {\rm exp}\Big(-\frac{1}{2}\sum_{{\bf r}={\bf r}_n}\Big[ \bar{\Psi}({\bf r}){\cal G}^{-1}(-i\partial_{\bf r}  \nonumber\\&& - {\bf A}({\bf r}))\Psi({\bf r})+ (h.c.)\Big]\Big)\label{Z3}
\end{eqnarray}
Here by $(h.c.)$ we denote the Hermitian conjugation, which is defined as follows. First of all, it relates the components of Grassmann variable $\Psi$ with the corresponding components of $\bar{\Psi}$. Besides, it inverses the ordering of operators and the variables $\bar{\Psi},\Psi$, and substitutes each operator by its Hermitian conjugated. For example, a conjugation of $\bar{\Psi} \hat{B} (i \partial_{r^{i_1}})...(i \partial_{r^{i_n}}) \Psi$ for a certain operator (in internal space) $\hat{B}$  is given by $\Big[(-i \partial_{r^{i_1}})...(-i \partial_{r^{i_n}}) \bar{\Psi} \Big]\hat{B}^+ \Psi$. As well as in continuum theory operators $\hat{p}_i - A_i({\bf r})$ and $\hat{p}_j - A_j({\bf r})$ do not commute for $i \ne j$. Therefore, we should point out the way of their ordering inside ${\cal G}^{-1}(-i\partial_{\bf r} - {\bf A}({\bf r}))$. We choose the following way for definiteness:  each product $p_{i_1} ... p_{i_n}$ in the expansion of ${\cal G}^{-1}$ is substituted by the symmetrized product $\frac{1}{n!}\sum_{\rm permutations} (\hat{p}_{i_1}-A_{i_1}) ... (\hat{p}_{i_n}-A_{i_n})$.
This method of introducing the gauge field to the lattice model differs from the more conventional ways, but it is manifestly gauge invariant, and it is obviously reduced to the conventional way the gauge field is to be introduced in the naive continuum limit. Therefore, it satisfies all requirements to be fulfilled by the introduction of the gauge field in lattice regularization of quantum field theory.

Now let us come back to momentum space:
\begin{eqnarray}
Z &=& \int D\bar{\Psi}D\Psi \, {\rm exp}\Big( - \frac{1}{2} \int_{\cal M} \frac{d^D {\bf p}}{|{\cal M}|} \Big[\bar{\Psi}({\bf p}){\cal Q}_{right}(i{\partial}_{\bf p},{\bf p})\Psi({\bf p}) \nonumber\\ &&+ \bar{\Psi}({\bf p}){\cal Q}_{left}(i{\partial}_{\bf p},{\bf p})\Psi({\bf p})  \Big]\Big)\label{Z4}
\end{eqnarray}
Here by ${\cal Q}_{right}$ we denote the function, that is constructed of ${\cal G}^{-1}$ as follows. We represent ${\cal G}^{-1}(-i\partial_{\bf r} - {\bf A}({\bf r}))$ as a series in powers of $-i\partial_{\bf r}$ and $ {\bf A}({\bf r})$ such that in each term  $ {\bf A}({\bf r})$ stand right to $-i\partial_{\bf r}$. For example, we represent $(-i \partial_{\bf r} - A({\bf r}))^2$ as $(-i \partial_{\bf r})^2 - 2(-i \partial_{\bf r}) {\bf A}({\bf r}) + {\bf A}^2({\bf r}) - i (\partial {\bf A})$. Next, we substitute the argument of ${\bf A}$ by $i \partial_{\bf p}$ and $-i\partial_{\bf r}$ by $\bf p$. Correspondingly,  ${\cal Q}_{left}$ is defined with the inverse ordering.

Let us introduce the following notation
\begin{equation}
\hat{\cal Q} = \frac{1}{2} \Big[{\cal Q}_{right}( i{\partial}^{}_{\bf p},{\bf p}) + {\cal Q}_{left}( i{\partial}^{}_{\bf p},{\bf p}) \Big]
\end{equation}
Since the commutators $[-i \partial_{r^i},r^j] = i \delta_i^j$ and $[p_i,i \partial_{p_j}] = i \delta_i^j$ are equal to each other, the actual expression for $\hat{\cal Q} $ is given by
\begin{equation}
\hat{\cal Q} = {\cal G}^{-1}({\bf p} - {\bf A}(i{\partial}^{}_{\bf p}))
\end{equation}
The Green function of our system in momentum space satisfies equation
\begin{equation}
\hat{\cal Q}(i \partial_{{\bf p}_1},{\bf p}_1)G({\bf p}_1,{\bf p}_2) = |{\cal M}| \delta^{(D)}({\bf p}_1-{\bf p}_2)\label{QGl}
\end{equation}

\subsection{Wigner transform in momentum space}
\label{SectWignLat}

According to the proposed above way to introduce the gauge field the Green function appears as a correlator
\begin{eqnarray}
G({\bf p}_1,{\bf p}_2)&=& \cor{-}\frac{1}{Z}\int D\bar{\Psi}D\Psi \,\bar{\Psi}({\bf p}_2)\Psi({\bf p}_1)\\ &&{\rm exp}\Big( - \int \frac{d^D {\bf p}}{|{\cal M}|} \bar{\Psi}({\bf p})\hat{Q}(i\partial_{\bf p},{\bf p})\Psi({\bf p}) \Big)\nonumber
\end{eqnarray}
It  obeys equation Eq. (\ref{QGl}).
Wigner transform \cite{Wigner} of the Green function may be defined as
\begin{equation}
 \tilde{G}({\bf R},{\bf p}) = \int \frac{d^D{\bf P}}{|{\cal M}|} e^{i {\bf P} {\bf R}} G({\bf p}+{\bf P}/2,{\bf p}-{\bf P}/2)\label{Wl}
\end{equation}
In terms of the Green function in coordinate space this Green function is expressed as:
\begin{equation}
 \tilde{G}({\bf R},{\bf p}) = \sum_{{\bf r}={\bf r}_n} e^{-i {\bf p} {\bf r}} G({\bf R}+{\bf r}/2,{\bf R}-{\bf r}/2)\label{Wl2}
\end{equation}
which is the direct analogue of Eq. (\ref{W}). In Appendix B we prove, that this Green function obeys the same equation as the one of the continuum theory:
\begin{eqnarray}
1 &=& {\cal Q}({\bf R},{\bf p})*\tilde G({\bf R},{\bf p})\nonumber\\ && \equiv {\cal Q}({\bf R},{\bf p})e^{\frac{i}{2}(\overleftarrow{\partial}_{\bf R}\overrightarrow{\partial}_{\bf p} - \overleftarrow{\partial}_{\bf p}\overrightarrow{\partial}_{\bf R})}\tilde G({\bf R},{\bf p})   \label{id}
\end{eqnarray}
As well as in continuum case the Weyl symbol of operator $\hat{\cal Q}$ is given by function $\cal Q$ that depends on the real numbers rather than on the operators. As it is explained in Appendix B, if $\hat{\cal Q}$ has the form of a function ${\cal G}^{-1}$ of the combination $({\bf p} - {\bf A}(\hat{\bf r}))$ with a gauge potential ${\bf A}(\hat{\bf r})$, i.e.
\begin{equation}
\hat{\cal Q}({\bf r},\hat{\bf p}) = {\cal G}^{-1}({\bf p} - {\bf A}(i\partial_{\bf p}))\label{Q}
\end{equation}
then we have
\begin{equation}
{\cal Q}({\bf r},{\bf p}) = {\cal G}^{-1}({\bf p} - {\bf A}({\bf r})) + O([\partial_i A_j]^2)\label{QF}
\end{equation}
Here $O([\partial_i A_j]^2)$ does not contain terms independent of the derivatives of $\bf A$ and the terms linear in those derivatives, i.e. it is higher order in derivatives. In certain particular cases the restrictions on the term $O(([\partial_i A_j]^2)$ may be more strong, or it may even vanish at all \cite{berezin}.

\subsection{Linear response of electric current to the strength of external gauge field.}
\label{SectLinLat}

In our lattice formalism the derivative expansion for the Wigner transform of the Green function is still given by Eq. (\ref{Gexp}), where
$\tilde G^{(0)}({\bf R},{\bf p})  = {\cal G}({\bf p}-{\bf A}({\bf R}))$.
Suppose, that we modified the external gauge field as ${\bf A} \rightarrow {\bf A} + \delta {\bf A}$. The response to this extra contribution to gauge potential gives electric current. Let us calculate this response basing on the description of the system given by Eq. (\ref{Z4}):
\begin{widetext}
\begin{eqnarray}
{\delta} \, {\rm log}\, Z&=& -\frac{1}{Z} \int D\bar{\Psi}D\Psi \, {\rm exp}\Big( - \int_{\cal M} \frac{d^D {\bf p}}{|{\cal M}|} \bar{\Psi}({\bf p})\hat{\cal Q}(i{\partial}_{\bf p},{\bf p})\Psi({\bf p}) \Big) \, \int_{\cal M} \frac{d^D {\bf p}}{|{\cal M}|} \bar{\Psi}({\bf p})\Big[\delta \hat{\cal Q}(i{\partial}_{\bf p},{\bf p})\Big]\Psi({\bf p}) \nonumber\\ & = & \cor{+}\int_{\cal M} \frac{d^D {\bf p}}{|{\cal M}|} \, {\rm Tr} \, \Big[ \delta \hat{\cal Q}(i{\partial}_{{\bf p}_1},{\bf p}_1)\Big]G({\bf p}_1,{\bf p}_2)\Big|_{{\bf p}_1 = {\bf p}_2 = {\bf p}}\nonumber\\ & = & \cor{+}\sum_{{\bf R}={\bf R}_n}\int_{\cal M} \frac{d^D {\bf p}}{|{\cal M}|} \,  {\rm Tr} \, \Big[ \delta \hat{\cal Q}(i{\partial}_{{\bf P}}+i{\partial}_{{\bf p}}/2 ,{\bf p}+{\bf P}/2)\Big]\, e^{-i {\bf P}{\bf R}} \tilde{G}({\bf R},{\bf p})\Big|_{{\bf P} = 0}  \label{j4}
\end{eqnarray}
\end{widetext}
In Appendix B we introduce function ${\cal Q}({\bf r},{\bf p})$ of real - valued arguments entering Eq. (\ref{id}). Notice, that $2{\bf p}$ and ${\bf P}$ enter the expression inside $\hat{\cal Q}$ in a symmetric way. This allows to use Eq. (\ref{corrl}). The form of Eq. (\ref{j4}) demonstrates, that the above expression for the electric current may also be written through the function $\cal Q$:
\begin{eqnarray}
{\delta} \, {\rm log}\, Z &=& \cor{+}\sum_{{\bf R}={\bf R}_n}\int_{\cal M} \frac{d^D {\bf p}}{|{\cal M}|} \,  {\rm Tr} \, \Big[ \delta {\cal Q}(i\overrightarrow{\partial}_{{\bf P}}-i\overleftarrow{\partial}_{{\bf p}}/2 ,{\bf p}+{\bf P}/2)\Big]\, \nonumber\\&&e^{-i {\bf P}{\bf R}} \tilde{G}({\bf R},{\bf p})\Big|_{{\bf P} = 0}\nonumber\\&=& \cor{+}\sum_{{\bf R}={\bf R}_n}\int_{\cal M} \frac{d^D {\bf p}}{|{\cal M}|} \,  {\rm Tr} \, \Big[ \delta {\cal Q}({\bf R} ,{\bf p}+{\bf P}/2)\Big]\, \nonumber\\&&e^{-i {\bf P}{\bf R}} \tilde{G}({\bf R},{\bf p})\Big|_{{\bf P} = 0}   \label{j428}
\end{eqnarray}
According to the notations of Appendix B the arrows above the derivatives mean, that those derivatives act only outside of ${\cal Q}$, and do not act on the arguments of $\cal Q$, i.e. $\overleftarrow\partial_{\bf p}$ acts on the function equal to $1$ standing left to the function $\cal Q$ while $\overrightarrow\partial_{\bf P}$ acts on the exponent $e^{-i {\bf P}{\bf R}} $.

As a result of the above mentioned manipulations we come to the following simple expression for the electric current per unit volume of coordinate space, which follows from the relation $\delta \, {\rm log}\, Z = \sum_{{\bf R}={\bf R}_n}{\bf j}^k({\bf R}) \delta A_k({\bf R})|{\cal V}|$:
\begin{eqnarray}
j^k({\bf R}) &=&\cor{-} \int_{\cal M} \frac{d^D {\bf p}}{|{\cal V}||{\cal M}|} \,  {\rm Tr} \, \tilde{G}({\bf R},{\bf p}) \frac{\partial}{\partial p_k}\Big[\tilde{G}^{(0)}({\bf R},{\bf p})\Big]^{-1}\label{j423}
\end{eqnarray}
Here by $|{\cal V}|$ we denote the volume of the unit cell understood as the ratio of the total volume of the system to the number of lattice points at which the field $\Psi$ is defined. For the ordinary hypercubic lattice the product of the two volumes is obviously  equal to $(2\pi)^D$. One might think, that for the lattices of more complicated symmetry the product of the momentum space volume and the defined above volume of the lattice cell may differ from this expression. Nevertheless, this is not so, and in general case the given product is always equal to $(2 \pi)^D$ exactly \footnote{For the purpose of illustration let us consider the 2D lattice of graphene \cite{Katsbook}.
In coordinate space we should take the hexagons that are formed by the atoms of sublattice $A$ (or $B$) because the resulting two component spinor is composed of the variables incident at the two sublattices. The resulting unit cell of the lattice is the hexagon surrounding each atom of the sublattice $A$. The length of its size is $a$, where $a$ is the distance between the adjacent $A$ and $B$ atoms. The area of the hexagon is equal to $|{\cal V}| = \frac{3\sqrt{3}}{2}\,a^2$. The Brillouin zone has also the form of the hexagon with the side length $\frac{2}{3\sqrt{3}}\,\frac{2\pi}{a}$. Its volume is $|{\cal M}| = \frac{2}{3\sqrt{3}}\,\frac{(2\pi)^2}{a^2}$. One can see, that the product is given by $(2\pi)^2$ as it should.}. In general case of an arbitrary crystal the direct proof is rather complicated. However, the result for the product of the two volumes may be found from the simple field theoretical correspondence: the limit of the microscopic model described by the effective low energy theory should correspond to the product of the two volumes equal to $(2\pi)^D$. Notice, that the construction of the unit cell in the original lattice should be performed with care. One has to count only those sites of the original crystal lattice, at which the dynamical variables of the model described by Eq. (\ref{Z3}) are incident. (This was illustrated above by the case of graphene, where we surrounded by this unit cell only the $A$ (or $B$) - atoms.)
Thus Eq. (\ref{j423}) coincides with the continuum expression Eq. (\ref{j14}).

Let us apply the gradient expansion to Eq. (\ref{j423}). It results in the following expression for the electric current:
\begin{eqnarray}
j^{k}({\bf R}) &=& j^{(0)k}({\bf R}) + j^{(1)k}({\bf R}) + ...\nonumber\\
j^{(0)k}({\bf R})  & = & \cor{-}\int \frac{d^D {\bf p}}{(2\pi)^D}\,  {\rm Tr}\, \tilde G^{(0)}({\bf R},{\bf p})\frac{\partial  \Big[\tilde G^{(0)}({\bf R},{\bf p})\Big]^{-1}}{\partial p_k}\label{j142}
\end{eqnarray}
Notice, that the second row of this expression represents the topological invariant as long as we deal with the system with regular Green functions, which do not have poles or zeros, i.e. this expression is unchanged while we are continuously deforming the Green function. We will not need this expression below since it does not contain the linear response to the external field strength.

In the $3+1$ D systems the contribution to this current originated from $\tilde{G}^{(1)}$ is given by
\begin{eqnarray}
j^{(1)k}({\bf R})  &= & \cor{-} \frac{1}{4\pi^2}\epsilon^{ijkl} {\cal M}_{l} A_{ij} ({\bf R}), \label{calM}\\
{\cal M}_l &=& \int_{} \,{\rm Tr}\, \nu_{l} \,d^4p \label{Ml} \\ \nu_{l} & = &  - \frac{i}{3!\,8\pi^2}\,\epsilon_{ijkl}\, \Big[  {\cal G} \frac{\partial {\cal G}^{-1}}{\partial p_i} \frac{\partial  {\cal G}}{\partial p_j} \frac{\partial  {\cal G}^{-1}}{\partial p_k} \Big]  \label{nuG}
\end{eqnarray}

In the linear response theory we should substitute ${\bf A}=0$ into the expression for ${\cal M}_l$. Therefore, in Eq. (\ref{nuG}) we substitute $\cal G$ instead of $\tilde{G}^{(0)}$. Further we will be interested in the component ${\cal M}_4$, which is the topological invariant, i.e. it is robust to any variations of the Green function $\tilde{G}$ as long as the singularities are not encountered (for the proof see Appendix C).

In the similar way for the $2+1$ D systems we get
\begin{eqnarray}
j^{(1)k}({\bf R})  &= & \cor{-} \frac{1}{2\pi}\epsilon^{ijk} \tilde{\cal N}_3 A_{ij} ({\bf R}), \label{calM2D}
\end{eqnarray}
where the topological invariant (denoted by $\tilde{\cal N}_3$ according to the classification of \cite{Volovik2003}) is to be calculated for the original system with vanishing background gauge field:
\begin{eqnarray}
\tilde{\cal N}_3 &=&  \frac{1}{24 \pi^2} {\rm Tr}\, \int_{} {\cal G}^{-1} d {\cal G} \wedge d {\cal G}^{-1} \wedge d {\cal G}\label{N3A}
\end{eqnarray}
Notice, that the above expression for ${\cal M}_l$ is the direct 4D generalization of the invariant $\tilde{\cal N}_3$. The proof that $\tilde{\cal N}_3$ is the topological invariant also follows from Appendix C.

It is worth mentioning, that in our calculations we use the Wigner transformation of the Green function without the parallel transporter factor. The Green functions with such a factor would be gauge invariant, and they are used, for example, in one of the methods of the chiral anomaly calculation \cite{anomaly_U}.  In this method the chiral current is expressed through the limit of the two point Green function $G(x,y)$ when $x \rightarrow y$, in which the parallel transporter between $x$ and $y$ is inserted into the definition of $G$. Such a Green function is manifestly gauge invariant. Although in the limit $x \rightarrow  y$ the mentioned parallel transporter tends to unity, its appearance is important when the derivative of the chiral current (i.e. anomaly) is calculated because in such a calculation the derivative over $(x+y)/2$ is taken first in order to have the finite expressions, and the limit $x \rightarrow y$ is taken at the end of the calculation. Thus, in such calculation of chiral anomaly the parallel transporter factor in the Green function is the tool necessary for the ultraviolet regularization that keeps gauge invariance.
This is in contrast to the expressions considered in the present paper. First of all, we do not calculate the derivative of the current. Therefore, even in Sect. \ref{SectCont}, where the continuum theory is discussed, we do not need such a factor in the Green function:  Eq.  (\ref{j14}) is manifestly gauge invariant. Moreover, the consideration of the present section deals with the lattice – regularization of theory, which is gauge invariant by construction. Indeed we use here the ordinary fermionic Green function, which is not gauge invariant, and depends on the external gauge field. The latter may be considered in a certain gauge, and with respect to the gauge transformations the status of our calculations is the same as the status of the perturbative calculations in quantum field theory performed in a certain fixed gauge. At the same time our final results of Eqs. (\ref{calM}) and(\ref{calM2D}) are gauge invariant by construction: they are obtained as variational derivatives with respect to the external gauge field of the manifestly gauge invariant lattice effective action Eq. (\ref{j4}).

\subsection{Applications to the $2+1$ D quantum Hall effect}
\label{SectHall}

In this section we demonstrate how the technique developed in the previous sections allows to reproduce the well - known expression for the Hall current in the gapped systems. Let us consider the $2+1 $ D model with the gapped fermions.
In the presence of external electric field ${\bf E} = (E_1,E_2)$ we substitute $A_{4k} = -i E_k$ into Eq. (\ref{calM2D}). This results in the following expression for the Hall current
\begin{equation}
{j}^k_{Hall} = \frac{1}{2\pi}\,\tilde{\cal N}_3\,\epsilon^{ki}E_i,\label{HALLj}
\end{equation}

Thus the well - known expression for the $2+1$ D QHE is reproduced (see Eqs. (11.1) and (21.12) of \cite{Volovik2003}). For one of the previous  derivations of this result see, for example, \cite{Volovik1988}.  The following remark is in order. In the real systems of finite sizes the total current is still given by this expression integrated over the direction of electric field, but the local current is concentrated close to the boundary. Our analysis based on the consideration of the systems of infinite volume does not allow to distinguish this inhomogeneity of current in coordinate space.

\section{Bulk chiral magnetic current.}
\label{SectCME}

\subsection{Chiral chemical potential and the Green function}
\label{Sectmu5}

Now let us concentrate on the $3+1$ D systems. We consider the situation, when vector gauge field $A_k({\bf R})$ has the nonzero components with $k=1,2,3$ that do not depend on (imaginary) time.
The conventional expression for the CME reads
\begin{equation}
j^{k}_{CME} = \frac{\mu_5}{4\pi^2}\,\xi_{CME}\,\epsilon^{ijk4}\, A_{ij}\label{jCSE}
\end{equation}
where $\xi_{CME}$ is integer number while $\mu_5$ is the chiral chemical potential\footnote{As it was mentioned above, our conventions of notations assume, that the field strength absorbs the elementary charge $e$, i.e. the Green function $\cal G$ contains the combination ${\bf p} -{\bf A}$ (where $\bf p$ is momentum) instead of the conventional ${\bf p}-e {\bf A}$ . Besides, the electric current $j$ is defined in the units of $e$, which eliminates the second factor $e$ from Eq. (\ref{jCSE}). Therefore, the conventional expression for the CME differs from Eq. (\ref{jCSE}) by the factor $e^2$.}. Such an expression should follow  from Eq. (\ref{calM}): we need to substitute ${\bf A}=0$ into Eq. (\ref{calM}) in the linear response approximation. Then one might expect that ${\cal M}_4 = \mu_5 \xi_{CME}$. However, below we will demonstrate, that $\xi_{CME}$ calculated in this way vanishes identically (for the reasonable choice of the way the chiral chemical potential is introduced) in the certain cases of interest.

There may exist many different definitions of $\mu_5$. Possibly, the most straightforward way is to consider the following expression for the fermion Green function:
 \begin{equation}
 {\cal G}^{}({\bf p}) = \Big(\sum_{k}\gamma^{k} g_{k}({\bf p}) + i\gamma^4 \gamma^5 \mu_5 - i m({\bf p})\Big)^{-1}\label{G2}
 \end{equation}
In the limit of vanishing chiral chemical potential it is reduced to
 \begin{equation}
 {\cal G}({\bf p})\Big|_{\mu_5=0} = \Big(\sum_{k}\gamma^{k} g_{k}({\bf p}) - i m({\bf p})\Big)^{-1}\label{G1}
 \end{equation}
 where $\gamma^k$ are Euclidean Dirac matrices while $g_k({\bf p})$ and $m({\bf p})$ are the real - valued functions, $k = 1,2,3,4$.
Here we define $\gamma^5$  in chiral representation as ${\rm diag}(1,1,-1,-1)$. It can be easily seen, that the consideration of the previous sections may be applied to the Green function, which has this form for nonzero value of $\mu_5$. Therefore, we may substitute ${\cal G}$ of Eq. (\ref{G2}) into Eq. (\ref{calM}) instead of $\tilde{G}({\bf R},{\bf p})$ while dealing with the linear response to the external magnetic field.

We are considering the theory with compact momentum space that can be represented as $S^1 \otimes \Omega$, where $\Omega$ is the compact $3D$ Brillouin zone. First, we assume, that with vanishing $\mu_5$ the Green functions do not have zeros or poles (at the real values of momenta), which means, that the fermions are massive. However, at the end of the calculations the limit of vanishing mass may always be considered.

Depending on the details of the given system the finite chiral chemical potential may induce the appearance of the poles of the Green function. For the Green function of the form of Eq. (\ref{G2}) the poles of the Green function appear as the zeros of ${\rm det} \, {\cal G}^{-1}_{}({\bf p}) \Big[{\cal G}^{-1}_{}({\bf p})\Big]^+$. The latter zeros are found as the solutions of the following equation
\begin{equation}
g^2_4({\bf p}) + \Big(\mu_5 \pm \sqrt{g_1^2({\bf p})+g_2^2({\bf p})+g_3^2({\bf p})}\Big)^2 + m^2({\bf p}) = 0\label{zero}
\end{equation}
Below we will analyse separately the electric current of Eq. (\ref{calM}) for the systems with and without poles of the Green function at the nonzero values of $\mu_5$.

\subsection{Massive fermions. The conventional case: nonzero $\mu_5$ does not induce the poles of $\cal G$.}
\label{Sectmu51}

For the massive fermions we consider as the conventional case the situation, when the nonzero $\mu_5$ does not induce the appearance of poles of the Green function. Let us explain this situation first by the particular example of the free lattice Wilson fermions with
\begin{equation}
g_k({\bf p}) = {\rm sin}\,
p_k, \quad m({\bf p}) = m^{(0)} +
\sum_{a=1,2,3,4} (1 - {\rm cos}\, p_a)\label{Wilson}
\end{equation}
Let us chose the conventional region of the values of parameter $m^{(0)} >0$. It corresponds to the vanishing value of the topological invariant $\tilde{\cal N}_5$ (see \cite{VZ2012}). In this case function $m({\bf p})$ never equals to zero. Therefore, the nonzero chiral chemical potential cannot cause the appearance of poles of the Green function.

 Since the lattice model with Wilson fermions  with $m^{(0)}$ is the typical (and in fact, the most popular) lattice regularization, we feel this appropriate to refer to the situation, when the chiral chemical potential does not cause poles of the Green function, as to the conventional case. It is realized, for example, for any Green function of the form of Eq. (\ref{G2}), such that there is no such value of ${\bf p}\in {\cal M}$, for which both $m({\bf p})$ and $g_4({\bf p})$ vanish.

This is in contrast to the case of the ordinary chemical potential added to the same system, in which the poles of the Green function may appear if the chemical potential exceeds the gap.

As it was mentioned above, ${\cal M}_4$ is topological invariant, i.e. it is robust to any variations of the Green function as long as the singularities are not encountered (for the proof see Appendix C). The introduction of chiral chemical potential is the particular case of such a variation. Therefore, ${\cal M}_4$ does not depend on $\mu_5$. Actually, for the Green function of the form of Eq. (\ref{G1}) ${\cal M}_4 = 0$, which may be checked by direct calculation:
\begin{eqnarray}
 {\cal M}_{4} &=& - \frac{i}{2}\int dp^4 \tilde{\cal N}_3(p^4),\label{calM40}\\
 \tilde{\cal N}_3(p^4) & = &  \frac{1}{24 \pi^2}\epsilon_{ijk4} {\rm Tr}\,  \int_{\Omega} d^3 p \Big( {\cal G} \partial^i {\cal G}^{-1} \Big)\nonumber\\&&\Big( {\cal G}\partial^j {\cal G}^{-1} \Big)\Big({\cal G} \partial^k {\cal G}^{-1}\Big) \label{F4B0}
\end{eqnarray}
Let us define the new auxiliary gamma - matrices $\Gamma^k = i\gamma^5\gamma^k$ for $k=1,2,3,4$, and $\Gamma^5 = \gamma^5$. One can easily check, that in terms of these gamma - matrices we have
\begin{eqnarray}
\tilde{\cal N}_3(p^4) & = &  \frac{1}{24 \pi^2}\epsilon_{ijk4} {\rm Tr}\, \Gamma^a\Gamma^b\Gamma^c\Gamma^d  \nonumber\\&& \int_{\Omega} d^3 p   \frac{g_a}{g^2} \partial^i  g_b \partial^j  \Big( \frac{g_c}{g^2} \Big) \partial^k {g_d} \nonumber\\
& = &  \frac{1}{6 \pi^2}\epsilon_{ijk4} (\delta^{ab}\delta^{cd}-\delta^{ac}\delta^{bd}+\delta^{ad}\delta^{bc})  \nonumber\\&&\int_{\Omega} d^3 p   \frac{g_a \partial^i  g_b \Big(\partial^j g_c - g_c \partial^j {\rm log}\,{g^2} \Big) \partial^k g_d}{g^4}
\nonumber\\
& = &  \frac{1}{6 \pi^2}\epsilon_{ijk4} (\delta^{ab}\delta^{cd}+\delta^{ad}\delta^{bc})  \nonumber\\&&\int_{\Omega} d^3 p   \frac{g_a \partial^i  g_b \partial^j g_c  \partial^k g_d}{g^4}=0
\label{F4B001}
\end{eqnarray}
 where  $g = \sqrt{\sum_{k=1,2,3,4,5}g_k^2}$.
Next, using continuous deformation of the Green function we may bring it to the form of Eq. (\ref{G2}) with nonzero $\mu_5$. During this deformation the poles of the Green function do not appear, and the value of ${\cal M}_4$ remains equal to zero.

 However, we do not need here the particular form of the Green function. We only need that momentum space is compact and can be represented as $S^1 \otimes \Omega$, where $\Omega$ is the compact $3D$ Brillouin zone.
The absence of the dependence of electric current on chiral chemical potential means, that there is no CME as long as we deal with compact momentum space and regular Green function. The finite value of chiral chemical potential does not change the situation in the conventional case because as we mentioned above such a finite value cannot provide the Green function of the form of Eq. (\ref{G1}) with the pole.

\subsection{Massive fermions. The finite values of $\mu_5$ that cause the appearance of the Fermi lines.}
\label{Sectmu52}

In this subsection we consider the marginal situation, when function $m({\bf p})$  may have zeros while at $\mu_5=0$ there are still no poles of $\cal G$. This situation may be illustrated by the model with Wilson fermions Eq. (\ref{Wilson}) for the values of the parameter $m^{(0)}$ such that the topological invariant $\tilde{\cal N}_5$ is nonzero (for the details see \cite{Z2012}). For example, if $m^{(0)} \in (-2,0)$, then the zeros of the function ${m}({\bf p})$ form the $3$D hyper - surfaces in momentum space. When $m^{(0)} \rightarrow -2$ these hyper - surfaces form the pairs, which approach (from the different sides) the hypersurfaces that connect the four of the $16$ fermion doublers $p_k = n_k \pi$, (with $n_k = 0,1$ and $n_1+n_2+n_3+n_4 = 1$) and that satisfy \begin{equation}
{\rm cos}\, p_1 + {\rm cos}\, p_2 + {\rm cos}\, p_3 + {\rm cos}\, p_4 = 2\label{hyper}
\end{equation}
This lattice system describes four (rather than one) physical massive fermions in the continuum limit.
The solutions of equation
\begin{equation}
|\mu_5| = \sqrt{g_1^2({\bf p})+g_2^2({\bf p})+g_3^2({\bf p})}\label{mu50}
\end{equation}
form the closed tubes extended in $p_4$ direction that enclose the positions of the fermion doublers $p_k = n_k \pi$ with $n_k = 0,1$ ($k=1,2,3$). We are interested in the sections of these tubes (being the $2D$ closed surfaces) at the values $p_4 = 0,\pi$, for which $g_4({\bf p})=0$, where the poles of the Green function may occur.  For the sufficiently small values of $\mu_5$ those sections do not intersect the positions of the zeros of $m({\bf p})$. In that case the poles of the Green function do not appear. However, for the sufficiently large values of $\mu_5$ the mentioned two types of surfaces intersect each other. The poles of the Green function appear along the closed Fermi lines in momentum space. Recall, that in case of the ordinary chemical potential we would deal with the Fermi surfaces, which have dimension $2$.

The value of ${\cal M}_4$ is given by Eq. (\ref{F4B0}).
For $p_4 \ne 0,\pi$ the quantity $\tilde{\cal N}_3(p_4)$ is the topological invariant (see Appendix C), and it has the same value when $\cal G$ is deformed smoothly. For example, we may perform the deformation, which brings $\mu_5$ to zero. For $\mu_5=0$ we have $\tilde{\cal N}_3(p_4) = 0$ because of the properties of the gamma - matrices. Thus, we come to the conclusion, that at $p_4\ne 0, \pi$ the value of $\tilde{\cal N}_3(p_4) $ vanishes at nonzero values of $\mu_5$. The integral over $p_4$ in Eq. (\ref{calM40}) may be regularized as  $\int_{-\pi+\epsilon}^{-\epsilon} + \int_\epsilon^{\pi-\epsilon}$. The limit $\epsilon \rightarrow 0$ should be taken at the end. With this regularization the value of ${\cal M}_4$ is equal to zero in the considered case.

In the similar way we may demonstrate that the value of ${\cal M}_4$ is equal to zero in the other systems that correspond to Eq. (\ref{G2}), when nonzero $\mu_5$ causes the appearance of  the Fermi lines provided that $g({\bf p})$ depends on $p_4$ only. This is the typical case for the non - interacting condensed matter systems and for the lattice regularization of the non - interacting gauge theory.

\subsection{Massless fermions}
\label{Sectmassless}

In principle, we may take the limit of vanishing mass of the final expression for the electric current in the above considered cases of massive lattice fermions, and in this limit the CME will be absent. However, we may also discuss from the very beginning the lattice massless fermions, which will be done in this section.

In this case the pole of the Green function appears and expression for ${\cal M}_4$ becomes ambiguous already for $\mu_5 = 0$. Let us consider separately the systems with vanishing fermion mass being the extensions of the systems  discussed in Sect. \ref{Sectmu51} and Sect. \ref{Sectmu52}.

\begin{enumerate}

\item{}

Let us set the fermion mass equal to zero in the conventional case of Sect. \ref{Sectmu51}. Again, let us start from the consideration of the Wilson fermions  Eq. (\ref{Wilson})  with the zero bare mass parameter $m^{(0)}=0$: in this example Eq. (\ref{zero}) does not have a solution for $\mu_5 \ne 0$. For $\mu_5 \ne 0$ the poles disappear, the expression for ${\cal M}_4$ becomes well - defined and independent of $\mu_5$.  Therefore, even for the gapless fermions our analysis gives the expression for the linear response of the electric current to the magnetic field that does not depend on chiral chemical potential as long as the latter is nonzero. This means, that the equilibrium CME is absent even for the fermions with zero mass.

For massless Wilson fermions with nonzero $\mu_5$ we may prove, that ${\cal M}_4 = 0$ as follows. At nonzero $\mu_5$ there are no poles of the Green function, and ${\cal M}_4$ is robust to the continuous transformations of the Green function, which do not give rise to such poles.  In particular, we may make $m^{(0)}$ nonzero in this way, which follows from Eq. (\ref{zero}). Finally, $\mu_5$ may be continuously brought to zero, in which case the calculation of ${\cal M}_4$ is trivial, and gives $0$.

 Following the same logic we may also prove, that ${\cal M}_4 = 0$ for the Green function with the form of Eq. (\ref{G2}) with nonzero chiral chemical potential $\mu_5$ under the following conditions:
1) The Green function does not have poles at the given value of $\mu_5$; 2) Function $m({\bf p})$ is either nonzero everywhere or may be brought to the form, when it does not have zeros, by the continuous deformation. During this deformation the common zeros of $m({\bf p})$ and $g_4({\bf p})$ should not cross the hyper - surface given by the solution of equation $$|\mu^{}_5|=\sqrt{g_1^2({\bf p})+g_2^2({\bf p})+g_3^2({\bf p})}$$

\item{}

Let us now discuss the modification with vanishing mass of the case considered in Sect. \ref{Sectmu52}. This pattern may again be considered using the example of Wilson fermions with $m^{(0)} = -2$. Function $m({\bf p})$ vanishes on the hypersurface, which is given by Eq. (\ref{hyper}). It connects the positions of those doublers $p_k = n_k \pi$ ($n_k = 0,1$), for which $n_1+n_2+n_3+n_4 = 1$. In this situation the nonzero value of $\mu_5$ does not eliminate the poles of the Green function: the surface given by Eq. (\ref{mu50}) intersects the $2D$ surface of the common zeros of $m({\bf p})$ and $g_4({\bf p})$. The manyfold of zeros of the Green function has dimension $1$ and represents the closed Fermi lines.  This is marginal case because as it was explained above, typically in the $3+1$ D systems the ordinary chemical potential causes the appearance of the Fermi surfaces of dimension $2$.

The value of ${\cal M}_4$ is still given by Eq. (\ref{calM40}), and the discussion given after this equation may be applied. Thus we obtain ${\cal M}_4 =0$. This consideration may be extended to the more general forms of functions $g_k$ and $m$ (such that the Fermi lines are produced by the nonzero value of $\mu_5$) if $g_4({\bf p})$ depends on $p_4$ only.

\end{enumerate}

Thus we have considered the wide class of lattice models. In our opinion the consideration of this class is enough to draw the conclusion, that the equilibrium CME is absent in the properly regularized quantum field theory. We also suppose, that our results indicate the absence of the CME in real Dirac semimetals. Nevertheless, at the present moment we do not exclude, that the CME may be possessed by certain marginal lattice models with unusual dependence of the Green function on momentum. Both existence of such hypothetical models and their relevance to physics remain unclear.

\section{Conclusions and discussion.}
\label{SectConcl}

In the present paper we use the methodology, which allows to reduce the consideration of the linear response of electric current (to external gauge field) to the discussion of momentum space topology.
We propose the original method to introduce the slow varying external gauge field to the lattice models. Although the proposed method looks unusual, it is manifestly gauge invariant, and it is obviously reduced in continuum limit to the conventional minimal connection of the fermionic theory with the gauge field. Therefore, it allows to introduce effectively the gauge field both into the lattice regularization of quantum field theory and to the models of the solid state physics. Since the proposed formalism in momentum space utilizes the pseudo - differential operators, i.e. the argument of the gauge field ${\bf A}(i\partial_{\bf p})$ is substituted by the differential operator, this formalism is not useful for the numerical simulations. However, it appears as a powerful tool for the analytical derivations.
The power of this methodology was demonstrated by the consideration of the $2+1$ D quantum Hall effect, where the conventional expression of the Hall conductivity through the topological invariant in momentum space is reproduced.

Further, we apply the same technique to the analysis of the equilibrium chiral magnetic effect. We show that the  corresponding current is also proportional to the momentum space topological invariant. We demonstrate, that this invariant is equal to zero for the wide class of systems with compact momentum space (that can be represented as $S^1 \otimes \Omega$, where $\Omega$ is the compact $3D$ Brillouin zone) and without poles or zeros of the Green function. This class includes the systems with nonzero chiral chemical potential described by certain lattice regularizations of quantum field theory, and certain solid state systems. The presence of the poles of the Green function at the real values of momenta and nonzero values of chiral chemical potential does not change this conclusion in several important  cases, which include the lattice regularizations with Wilson fermions and the similar models of Dirac semimetals.
Therefore, we conclude, that the properly regularized quantum field theory does not possess the equilibrium bulk chiral magnetic effect. Although we did not considered the general case of arbitrary Dirac semimetals, our results indicate, that there is no equilibrium  CME in those materials as well.

We calculated the response to the external magnetic field in the system of the non - interacting fermions. However, according to the general properties of the topological invariants, they cannot be changed by the continuous deformation of the system. Therefore, the mentioned above conclusion on the topological contribution to the CME remains the same if we turn on the self - interactions. The expression for the topological contribution to the considered current remains unchanged until the phase transition is encountered. Our considerations do not exclude that the self - interactions cause the non - topological contribution to electric current proportional to chiral chemical potential and external magnetic field. However, since such a contribution is not related to topology, it must be dissipative, which excludes its appearance because magnetic field cannot cause heat.

Our conclusion on the absence of bulk equilibrium CME current is in accordance with the recent lattice calculations of the CME by Buividovich and co - authors \cite{Buividovich:2015ara,Buividovich:2015ara,Buividovich:2014dha,Buividovich:2013hza}, it is also in accordance with the consideration of the particular model of Weyl semimetals \cite{nogo} and with the no - go Bloch theorem \cite{nogo2}.  (The formulations used in \cite{nogo2}, however, seem to the author rather distant from the setup of the present paper.) Besides, our conclusion is in line with the discussion of the CME using the continuous model with special boundary conditions in the direction of magnetic field \cite{Gorbar:2015wya}. However, the methodology presented in the present paper is rather general, and it allows to draw the conclusion on the absence of the equilibrium CME for the wide class of systems, which is not limited to the particular models considered in the mentioned above papers.

Our conclusions also do not contradict with the certain calculations made within the continuum relativistic field theory in the Pauli - Villars regularization. Namely, in \cite{Buividovich:2013hza} (see also Eq. (1.2) in \cite{Valgushev:2015pjn} and the earlier paper \cite{Hou:2011ze}) the CME current was calculated in this regularization for the inhomogeneous magnetic field. It was demonstrated, that the frequency dependence of the CME conductivity gives zero in the limit ${\rm lim}_{\vec{q}\to 0}{\rm lim}_{q_0 \to 0}$ (here $q = (q_0,\vec{q})$ is the four - momentum corresponding to the inhomogeneous magnetic field). It is also worth mentioning, that the conclusion on the absence of the CME in the properly defined quantum field theory was reached within the holographic pattern (see \cite{Rebhan,Rubakov} and references therein). It was noticed in \cite{Rubakov} that the notion of chiral chemical potential may be redefined in the way different from that of discussed in the present paper and in the mentioned above publications. Namely, it may be defined as the chemical potential corresponding to the conserved chiral charge. The latter is given by the sum of the naive, non - conserved chiral charge and the Chern - Simons form. The Chern - Simons term being multiplied by $\mu_5$  induces the current, which formally coincides with the naive CME expression. Therefore, in \cite{Rubakov} the conclusion was drawn, that with this modification of the notion of chiral chemical potential the CME is back\footnote{See also \cite{Gorsky}.}. At this point we would like to emphasise the essential difference between this modified understanding of the chiral magnetic effect and its original understanding, which assumes, that the chiral chemical potential is introduced according to Eq. (\ref{G2}). In the present paper we demonstrate that the CME understood in its original form is absent while that of \cite{Rubakov} is indeed back.

Notice, that our conclusion refers to the equilibrium states only. The contribution to the conductivity in the presence of both electric and magnetic fields \cite{Nielsen:1983rb,ZrTe5}) due to the chiral chemical potential induced by chiral anomaly may avoid the restrictions imposed on the CME by momentum space topology. This may be related to essentially non - equilibrium nature of this phenomenon. The notion of the chiral chemical potential generated by the interplay of chiral anomaly and the quasiparticle interactions with the change of chirality may differ from the notion of the chiral chemical potential of equilibrium theory. Actually, the given contribution to conductivity is to be described by the higher orders of the derivative expansion for the Wigner transform of the Green function just because it is expected to be proportional to the magnetic field squared. This could restore the CME (possibly, in a modified form). Besides, the experience of the quantum Hall effect prompts, that inclusion into consideration of boundaries may be important. These issues are, however, out of the scope of the present paper.

\section*{Acknowledgements}
The author is grateful to G.E. Volovik, who pointed out that the equilibrium CME in solids may contradict to Bloch theorem, and thus changed the point of view of the author on the subject of the present paper. The author also kindly acknowledges useful discussions with M.N.Chernodub. This work was supported by LE STUDIUM research
fellowship.

%The present work was partially supported by Russian Science
%Foundation Grant No 16-12-10059.

\section*{Appendix A. Wigner transform of the Green function}
\label{SectWigner}

Let us consider the $d+1 = D$ dimensional model with the Green function ${\cal G}({\bf r}_1,{\bf r}_2)$ that obeys equation
\begin{equation}
\hat{\cal Q}({\bf r}_1,-i \partial_{{\bf r}_1})G({\bf r}_1,{\bf r}_2) =  \delta^{(D)}({\bf r}_1-{\bf r}_2)\label{QG}
\end{equation}
for some Hermitian operator - valued function $\cal Q$. Let us apply Wigner decomposition
\begin{equation}
 \tilde{G}({\bf R},{\bf p}) = \int d^D{\bf r} e^{-i {\bf p} {\bf r}} G({\bf R}+{\bf r}/2,{\bf R}-{\bf r}/2)\label{W0}
\end{equation}
Below we will prove the following identity
\begin{equation}
 {\cal Q}({\bf R},{\bf p})e^{\frac{i}{2}(\overleftarrow{\partial}_{\bf R}\overrightarrow{\partial}_{\bf p} - \overleftarrow{\partial}_{\bf p}\overrightarrow{\partial}_{\bf R})}\tilde G({\bf R},{\bf p})  = 1 \label{idB0}
\end{equation}
Here the function $\cal Q$ depends on the real numbers rather than on the operators. $\cal Q$ is called the Weyl symbol of the operator $\hat{\cal Q}$ \cite{Weyl}. We determine relation between the function ${\cal Q}({\bf r},{\bf p})$ (of real - valued vectors ${\bf r}$ and ${\bf p}$) and the function $\hat{\cal Q}({\bf r}, \hat{\bf p})$ (of the operators ${\bf p}$ and $\hat{\bf p} = - i \partial_{\bf r}$) through the identity
\begin{eqnarray}
&& \int  d^D {\bf r}\, f({\bf r},{\bf R})\, {\cal Q}({\bf R}+\frac{\bf r}{2},i \overleftarrow{\partial}_{\bf r}-\frac{i}{2}\overrightarrow{\partial}_{\bf R})\, h({\bf r},{\bf R})\nonumber\\ && =  \int  d^D {\bf r}\, f({\bf r},{\bf R}) \hat{\cal Q}\Big({\bf R}+\frac{\bf r}{2},-i \frac{\partial}{\partial ({\bf R}+ \frac{\bf r}{2})}\Big) \, h({\bf r},{\bf R}) \label{corr}
\end{eqnarray}
which works for arbitrary functions $f({\bf r},{\bf R})$ and $h({\bf r},{\bf R})$ that decrease sufficiently fast at infinity. The important point concerning this expression is that the derivatives  $\overrightarrow{\partial}_{\bf R}$ and $\overleftarrow{\partial}_{\bf r}$ inside the arguments of $\cal Q$ act only outside of this function, i.e. $\overleftarrow{\partial}_{\bf r}$ acts on $f({\bf r},{\bf R})$ while $\overrightarrow{\partial}_{\bf R}$  acts on $h({\bf r},{\bf R})$. At the same time
the derivatives without arrows act as usual operators, i.e. not only right to the function $\hat{\cal Q}$, but inside it as well. Notice, that $\frac{\partial}{\partial ({\bf R}+ \frac{\bf r}{2})} = \partial_{\bf r} + \frac{1}{2}{\partial_{\bf R}}$.

The given correspondence looks rather complicated. However, it takes the simple form in certain particular cases.
For example, if $\hat{\cal Q} = (\hat{\bf p} - {\bf A}({\bf r}))^2 = \hat{\bf p}^2 + {\bf A}^2({\bf r}) + i \Big(\partial^k A_k({\bf r})\Big) - 2 {\bf A}({\bf r})\hat{\bf p}$, then  ${\cal Q} = {\bf p}^2 + {\bf A}^2({\bf r}) - 2 {\bf A}({\bf r}){\bf p}$. Besides, one can easily check, that if $\hat{\cal Q}$ has the form
\begin{equation}
\hat{\cal Q}({\bf r},\hat{\bf p}) = {\cal F}(\hat{\bf p} - {\bf A}({\bf r}))
\end{equation}
then we have
\begin{equation}
{\cal Q}({\bf r},\hat{\bf p}) = {\cal F}({\bf p} - {\bf A}({\bf r})) + O([\partial_i A_j]^2)\label{QF0}
\end{equation}
Here $O([\partial_i A_j]^2)$ may contain the terms with the second power of the derivatives of $\bf A$ and the terms higher order in derivatives. In order to prove Eq. (\ref{QF0}) it is necessary to consider the function ${\cal F}(\hat{\bf p} - {\bf A}({\bf r})) = \sum_{n}{\cal F}_{i_1...i_n}(-i\partial_{i_1} - A_{i_1}({\bf r}))...(-i\partial_{i_n} - A_{i_n}({\bf r}))$ as a series in powers of its arguments (${\cal F}_{i_1...i_n}$ are the Hermitian operator - valued coefficients), and apply the correspondence of Eq. (\ref{corr}) to each term. The details of the consideration are similar to that of Appendix B, where the Wigner transform in momentum space is discussed. Therefore, we do not represent them here and advise the reader to follow Appendix B.

 In order to prove Eq. (\ref{idB0}) let us substitute Eq. (\ref{W0}) into it.
Argument of the exponent in Eq. (\ref{idB0}) acts on $\cal Q$ as follows:
\begin{eqnarray}
1&=&\int d^D{\bf r}  {\cal Q}({\bf R}+\frac{i}{2}\overrightarrow{\partial}_{\bf p},{\bf p}-\frac{i}{2}\overrightarrow{\partial}_{\bf R}) \nonumber\\&& e^{-i {\bf p} {\bf r}} G({\bf R}+{\bf r}/2,{\bf R}-{\bf r}/2)
\end{eqnarray}
The important point concerning this expression is that the derivatives  $\overrightarrow{\partial}_{\bf R}$ and $\overrightarrow{\partial}_{\bf p}$ inside the arguments of $\cal Q$ act only outside of this function, i.e. on $e^{-i {\bf p} {\bf r}} G({\bf R}+{\bf r}/2,{\bf R}-{\bf r}/2)$ and do not act inside the function $\cal Q$, i.e. on $\bf p$ and $\bf R$ in its arguments.
This gives
\begin{eqnarray}
1&=&\int d^D{\bf r} e^{-i {\bf p} {\bf r}} {\cal Q}({\bf R}+\frac{\bf r}{2},i \overleftarrow{\partial}_{\bf r}-\frac{i}{2}\overrightarrow{\partial}_{\bf R})\\&&  G({\bf R}+{\bf r}/2,{\bf R}-{\bf r}/2)  \nonumber
\end{eqnarray}
Up to the boundary terms (which are assumed to be absent) we arrive at
\begin{equation}
\int d^D{\bf r} e^{-i {\bf p} {\bf r}} \hat{\cal Q}({\bf R}+\frac{\bf r}{2},-i {\partial}_{\bf r}-\frac{i}{2}{\partial}_{\bf R})   G({\bf R}+{\bf r}/2,{\bf R}-{\bf r}/2)  = 1\nonumber
\end{equation}
Now it is clear why we should order operators $\hat{\bf p}$ and $\bf r$ in $\hat{\cal Q}$ according to Eq. (\ref{corr}) in order to obtain function $\cal Q$.
Applying the inverse Wigner transform we finally arrive at Eq. (\ref{QG}).

Notice, that the Weyl symbol $\cal Q$ of the operator $\hat{\cal Q}$ may also be defined as \cite{Weyl,berezin} the Wigner transform of the matrix elements of $\hat{\cal Q}$:
\begin{eqnarray}
 {\cal Q}({\bf R},{\bf p}) &=& \int d^D {\bf x} d^D{\bf r} e^{-i {\bf p} {\bf r}} \delta({\bf R}-{\bf r}/2 - {\bf x}) \nonumber\\&& \hat{\cal Q}({\bf x},-i {\partial}_{\bf x})   \delta({\bf R}+{\bf r}/2 - {\bf x})\label{W000}
\end{eqnarray}

\section*{Appendix B. Wigner transform of the Green function (lattice version)}
\label{SectWignerl}

\cor{Here, as well as in the main text, we consider the case when the system contains weak inhomogeneity, which may be neglected at the distance of the order of the lattice spacing.} We consider the $d+1 = D$ dimensional model with the Green function ${\cal G}({\bf p}_1,{\bf p}_2)$ that obeys equation
\begin{equation}
\hat{\cal Q}(i \partial_{{\bf p}_1},{\bf p}_1)G({\bf p}_1,{\bf p}_2) = |{\cal M}| \delta^{(D)}({\bf p}_1-{\bf p}_2)\label{QGlA}
\end{equation}
for some Hermitian operator - valued function $\hat{\cal Q}$. Let us apply Wigner decomposition in momentum space
\begin{equation}
 \tilde{G}({\bf R},{\bf p}) = \int \frac{d^D{\bf P}}{|{\cal M}|} e^{i {\bf P} {\bf R}} G({\bf p}+{\bf P}/2,{\bf p}-{\bf P}/2)\label{WlA}
\end{equation}
We will prove identity
\begin{equation}
 {\cal Q}({\bf R},{\bf p})e^{\frac{i}{2}(\overleftarrow{\partial}_{\bf R}\overrightarrow{\partial}_{\bf p} - \overleftarrow{\partial}_{\bf p}\overrightarrow{\partial}_{\bf R})}\tilde G({\bf R},{\bf p})  = 1 \label{idBl}
\end{equation}
Weyl symbol $\cal Q$ of the operator $\hat{\cal Q}$ is the function of real numbers rather than the operators. Similar to the continuum case we determine relation between the function ${\cal Q}({\bf r},{\bf p})$ (of real - valued vectors ${\bf r}$ and ${\bf p}$) and the function $\hat{\cal Q}(\hat{\bf r}, {\bf p})$ (of the operators ${\bf p}$ and $\hat{\bf r} = i \partial_{\bf p}$) through the identity
\begin{widetext}
\begin{eqnarray}
&& \int d^D {\bf X} d^D {\bf Y}\, f({\bf X},{\bf Y})\, {\cal Q}(-i \overleftarrow{\partial}_{\bf Y}+i\overrightarrow{\partial}_{\bf X},{\bf X}/2+{\bf Y}/{2})\, h({\bf X},{\bf Y})\nonumber\\ && =  \int  d^D {\bf X}d^D {\bf Y}\, f({\bf X},{\bf Y}) \hat{\cal Q}\Big(i \partial_{\bf X}+ i\partial_{\bf Y},{\bf X}/2+ {\bf Y}/{2}\Big) \, h({\bf X},{\bf Y}) \label{corrl}
\end{eqnarray}
\end{widetext}
which works for arbitrary functions $f({\bf X},{\bf Y})$ and $h({\bf X},{\bf Y})$ defined on compact momentum space ${\bf X},{\bf Y}\in {\cal M}$. The important point concerning this expression is that the derivatives  $\overrightarrow{\partial}_{\bf X}$ and $\overleftarrow{\partial}_{\bf Y}$ inside the arguments of $\cal Q$ act only outside of this function, i.e. $\overleftarrow{\partial}_{\bf Y}$ acts on $f({\bf X},{\bf Y})$ while $\overrightarrow{\partial}_{\bf X}$  acts on $h({\bf X},{\bf Y})$. At the same time
the derivatives without arrows act as usual operators, i.e. not only right to the function $\hat{\cal Q}$, but inside it as well. Notice, that $\frac{\partial}{\partial ({\bf X}/2+ {\bf Y}/{2})} = \partial_{\bf Y} + {\partial_{\bf X}}$ and $\frac{\partial}{\partial ({\bf X}/2 - {\bf Y}/{2})} = \partial_{\bf X} - {\partial_{\bf Y}}$. Therefore, we may rewrite Eq. (\ref{corrl}) as
\begin{widetext}
\begin{eqnarray}
&& \int d^D {\bf X} d^D {\bf Y}\, f({\bf X},{\bf Y})\, {\cal Q}(-i \overleftarrow{\partial}_{\bf Y}+i\overrightarrow{\partial}_{\bf X},{\bf X}/2+{\bf Y}/{2})\, h({\bf X},{\bf Y})\nonumber\\ && = -2 \int d^D {\bf Q} d^D {\bf K}\, f({\bf Q}+{\bf K},{\bf Q}-{\bf K}) \hat{\cal Q}\Big(i \partial_{\bf Q},{\bf Q}\Big) \, h({\bf Q}+{\bf K},{\bf Q}-{\bf K}) \label{corrl2}
\end{eqnarray}
\end{widetext}

The given correspondence takes the simple form in certain particular cases.
For example, if $\hat{\cal Q} = ({\bf p} - {\bf A}(\hat{\bf r}))^2 = {\bf p}^2 + {\bf A}^2(\hat{\bf r}) + i \Big(\partial^k A_k(\hat{\bf r})\Big) - 2 {\bf A}(\hat{\bf r}){\bf p}$ (recall, that  $\hat{\bf r}$ is operator equal to $i\partial_{\bf p}$), then  ${\cal Q} = {\bf p}^2 + {\bf A}^2({\bf r}) - 2 {\bf A}({\bf r}){\bf p}$. Besides, one can easily check, that if $\hat{\cal Q}$ has the form
\begin{equation}
\hat{\cal Q}(\hat{\bf r},{\bf p}) = {\cal F}({\bf p} - {\bf A}(\hat{\bf r}))
\end{equation}
then we have
\begin{equation}
{\cal Q}({\bf r},{\bf p}) = {\cal F}({\bf p} - {\bf A}({\bf r})) + O([\partial_i A_j]^2)\label{QF}
\end{equation}
Here $O([\partial_i A_j]^2)$ may contain the terms with the second power of the derivatives of $\bf A$  and the terms higher order in derivatives. Let us prove Eq. (\ref{QF}). First of all, this is necessary to consider the function ${\cal F}({\bf p} - {\bf A}(\hat{\bf r})) = \sum_{n}{\cal F}_{i_1...i_n}(p_{i_1} - A_{i_1}(i\partial_{\bf p}))...(p_{i_n} - A_{i_n}(i\partial_{\bf p}))$ as a series in powers of its arguments (${\cal F}_{i_1...i_n}$ are Hermitian operators that do not depend on $\bf p$).  Operator $\hat{Q}$ is Hermitian, therefore, the kernel of the first row in Eq. (\ref{corrl2}) should also be Hermitian.
It may be represented as follows. Suppose, that function $\cal Q$ is expanded in powers of ${\bf Q}=({\bf X}+{\bf Y})/2$ and $-i \overleftarrow{\partial}_{\bf Y}+i\overrightarrow{\partial}_{\bf X}$ as follows
\begin{widetext}
\begin{equation}
{\cal Q}(-i \overleftarrow{\partial}_{\bf Y}+i\overrightarrow{\partial}_{\bf X},{\bf X}/2+{\bf Y}/{2}) = \sum q_{i_1...i_n;j_1...j_m;k_1...k_l}(-i \overleftarrow{\partial}_{Y_{i_1}})...(-i \overleftarrow{\partial}_{Y_{i_n}}) Q_{j_1}...Q_{j_m}(i\overrightarrow{\partial}_{X_{k_1}})(i\overrightarrow{\partial}_{X_{k_l}})\label{QQ}
\end{equation}
\end{widetext}
In this expression inside the first row of Eq. (\ref{corrl}) we may be substitute $-i\overleftarrow{\partial}_{Y_{i}}$ by  $i{\partial}_{Y_{i}}$ and $i\overrightarrow{\partial}_{X_{k}}$ by $i{\partial}_{X_{k}}$. Because the second row in Eq. (\ref{corrl}) is symmetric under the interchange of $\bf X$ and $\bf Y$, we have $q_{i_1...i_n;j_1...j_m;k_1...k_l}=q_{k_1...k_l;j_1...j_m;i_1...i_n}$. For the same reason Eq. (\ref{QQ}) is invariant under the interchange ${\bf X}\leftrightarrow {\bf Y}$. Then the change of $q_{...}$ by its Hermitian conjugate $q^+_{...}$ is equivalent to the Hermitian conjugation of the whole expression. This demonstrates that coefficients $q_{...}$ are Hermitian. Now let us suppose, that ${\cal Q}$ is linear in the derivative of ${\bf A}$. Algebraically the linear term appears as a product of a certain combination of ${\cal F}_{...}$ and the commutator $[{p}_k,{\bf A}(i\partial_{\bf p})] = - i (\partial_k {\bf A})$. Therefore, it would lead to the appearance of imaginary unity in the expression for $q_{...}$ as a combination of ${\cal F}_{...}$, which means that $q_{...}$ is not Hermitian. The contradiction proves the non - appearance of the terms linear in the derivatives of ${\bf A}$ in the expression for ${\cal Q}({\bf r},{\bf p})$.

 In order to prove Eq. (\ref{idBl}) let us substitute Eq. (\ref{WlA}) into it.
Argument of the exponent in Eq. (\ref{idBl}) acts on $\cal Q$ as follows:
\begin{eqnarray}
1&=&\int \frac{d^D{\bf P}}{|{\cal M}|}  {\cal Q}({\bf R}+\frac{i}{2}\overrightarrow{\partial}_{\bf p},{\bf p}-\frac{i}{2}\overrightarrow{\partial}_{\bf R}) \nonumber\\&& e^{i {\bf P} {\bf R}} G({\bf p}+{\bf P}/2,{\bf p}-{\bf P}/2)
\end{eqnarray}
In this expression the derivatives  $\overrightarrow{\partial}_{\bf p}$ and $\overrightarrow{\partial}_{\bf R}$ inside the arguments of $\cal Q$ act only outside of this function, i.e. on $e^{i {\bf P} {\bf R}} G({\bf p}+{\bf P}/2,{\bf p}-{\bf P}/2)$ and do not act inside the function $\cal Q$, i.e. on $\bf p$ and $\bf R$ in its arguments.
This gives
\begin{eqnarray}
1&=&\int \frac{d^D{\bf P}}{|{\cal M}|} e^{i {\bf P} {\bf R}} {\cal Q}(-i \overleftarrow{\partial}_{\bf P}+\frac{i}{2}\overrightarrow{\partial}_{\bf p},{\bf p}+\frac{\bf P}{2})\\  && G({\bf p}+{\bf P}/2,{\bf p}-{\bf P}/2)  \nonumber
\end{eqnarray}
Because of the absence of boundary of $\cal M$ we arrive at
\begin{equation}
\int \frac{d^D{\bf P}}{|{\cal M}|} e^{i {\bf P} {\bf R}} \hat{\cal Q}(i {\partial}_{\bf P}+\frac{i}{2}{\partial}_{\bf p},{\bf p}+\frac{\bf P}{2})   G({\bf p}+{\bf P}/2,{\bf p}-{\bf P}/2)  = 1\nonumber
\end{equation}
where we applied Eq. (\ref{corrl}). Now it is clear why we should order operators ${\bf p}$ and $\bf{\bf r}$ in $\hat{\cal Q}$ according to Eq. (\ref{corrl}) in order to obtain function $\cal Q$.
Taking into account Eq. (\ref{corrl2}) and applying the inverse Wigner transform we finally arrive at Eq. (\ref{QGlA}).

Finally, let us notice, that the Weyl
symbol $\cal Q$ of the operator $\hat{\cal Q}$ may also be defined following that of Appendix A (see also \cite{Weyl,berezin}) as
\begin{eqnarray}
 {\cal Q}({\bf R},{\bf p}) &=& \int {d^D {\bf K}} {d^D{\bf P}} e^{i {\bf P} {\bf R}} \delta({\bf p}-{\bf P}/2 - {\bf K}) \nonumber\\&& \hat{\cal Q}(i {\partial}_{\bf K},{\bf K})   \delta({\bf p}+{\bf P}/2 - {\bf K})\label{W000}
\end{eqnarray}

\section*{Appendix C. Topological invariant responsible for the linear response of electric current to magnetic field}

\label{SectN34}

In the main text we encountered Eq. (\ref{Ml}) for the coefficient entering the linear response of electric current to external magnetic field. If momentum space $\cal M$ has the form of the product $S^1\otimes \Omega$, where $\Omega$ is the 3D Brillouin zone, while $S^1$ is the circle of the values of $p_4$, then for $l=4$ we may rewrite this quantity as follows:
\begin{eqnarray}
 {\cal M}_{4} &=& - \frac{i}{2}\int dp^4 \tilde{\cal N}_3(p^4),\nonumber\\
 \tilde{\cal N}_3(p^4) & = &  \frac{1}{24 \pi^2}\epsilon_{ijk4} {\rm Tr}\,  \int_{\Omega} d^3 p \Big( {\cal G} \partial^i {\cal G}^{-1} \Big)\nonumber\\&&\Big( {\cal G}\partial^j {\cal G}^{-1} \Big)\Big({\cal G} \partial^k {\cal G}^{-1}\Big) \label{F4B}
\end{eqnarray}
Here for the fixed value of $p^4$ we encounter the expression for the topological invariant in the 3D Brillouin zone. Green function $\cal G$ should be considered here as the function of the 3 arguments $p^1,p^2,p^3$ while $p^4$ is to be considered as a parameter.

Notice, that for the Green function of the form of Eq. (\ref{G1}) the value of $\tilde{\cal N}_3(p^4)$ is equal to zero. At the same time for the Green function of general form this invariant may be nonzero. This explains the quantization of Hall conductivity as has been explained in Sect. \ref{SectHall}.  One might naively think, that the deviation of the Green function from the form of Eq. (\ref{G1}) - say, of the form of Eq. (\ref{G2}) may change the expressions for $\tilde{\cal N}_3$ and ${\cal M}_4$. Below we will demonstrate, that this does not occur as long as we deal with the compact Brillouin zone and regular Green functions.

Let us consider arbitrary variation of the Green function:
${\cal G} \rightarrow {\cal G} + \delta {\cal G}$. Then expression
for $\tilde{\cal N}_3$ is changed as follows:
\begin{eqnarray}
  \delta \tilde{\cal N}_3  & = &-
  \frac{3 }{24  \pi^2} \int_{} {\rm Tr} \Big( ([\delta {\cal G}]
  d  {\cal G}^{-1}+{\cal G}
  d  [\delta {\cal G}^{-1}])\wedge {\cal G}
  d  {\cal G}^{-1} \nonumber\\&& \wedge {\cal G}
  d  {\cal G}^{-1}\Big)\nonumber\\&=&
\frac{3}{24  \pi^2} \int_{} d \, {\rm Tr} \left(
([\delta {\cal G}^{-1}] {\cal G} )
  d  {\cal G}^{-1}\wedge
  d  {\cal G}\right) = 0 \label{dNP}
\end{eqnarray}
Thus  we proved that $\tilde{\cal N}_3$ is the topological invariant.

%%%%%%%%%%%%%%%%%%%%%%%%%%%%%%%%%%%%%%%%%%%%%%%%%%%%%%%%%%%%%%%%%%%%%%%%%%%%%%%%%%%%%%%%%%%%%%%%%%%%%%%%%%%%%%%%%%%%%%%%%%%%%%%%%%%%%%%%%%%%%%


\begin{thebibliography}{95}


\bibitem{Vilenkin}  A. Vilenkin, Phys. Rev. D 22, 3080 (1980)

\bibitem{CME}
 K. Fukushima, D.E. Kharzeev, H.J. Warringa, Phys.Rev.D 78:074033,2008

\bibitem{Kharzeev:2013ffa}
  D.~E.~Kharzeev,
  ``The Chiral Magnetic Effect and Anomaly-Induced Transport,''
  Prog.\ Part.\ Nucl.\ Phys.\  {\bf 75} (2014) 133
  doi:10.1016/j.ppnp.2014.01.002
  [arXiv:1312.3348 [hep-ph]].
  %%CITATION = doi:10.1016/j.ppnp.2014.01.002;%%
  %65 citations counted in INSPIRE as of 10 févr. 2016

\bibitem{SonYamamoto2012}
D.T.  Son and N. Yamamoto,
{\sl ''Berry curvature, triangle anomalies, and chiral magnetic effect in Fermi liquids''}, Phys.Rev.Lett.109:181602,2012,
 arXiv:1203.2697.

\bibitem{Nielsen:1983rb}
  H.~B.~Nielsen and M.~Ninomiya,
  ``Adler-bell-jackiw Anomaly And Weyl Fermions In Crystal,''
  Phys.\ Lett.\ B {\bf 130} (1983) 389.
  doi:10.1016/0370-2693(83)91529-0
  %%CITATION = doi:10.1016/0370-2693(83)91529-0;%%
  %211 citations counted in INSPIRE as of 02 févr. 2016

\bibitem{Kharzeev:2009pj}
  D.~E.~Kharzeev and H.~J.~Warringa,
  ``Chiral Magnetic conductivity,''
  Phys.\ Rev.\ D {\bf 80} (2009) 034028
  doi:10.1103/PhysRevD.80.034028
  [arXiv:0907.5007 [hep-ph]].
  %%CITATION = doi:10.1103/PhysRevD.80.034028;%%
  %138 citations counted in INSPIRE as of 02 févr. 2016


\bibitem{ZrTe5}
Q. Li, D. E. Kharzeev, C. Zhang, Y. Huang, I. Pletikosic, A. V. Fedorov, R. D. Zhong, J. A. Schneeloch, G. D. Gu, and T. Valla,
arXiv:1412.6543.






%\cite{Landsteiner:2012kd}
\bibitem{Landsteiner:2012kd}
  K.~Landsteiner, E.~Megias and F.~Pena-Benitez,
  ``Anomalous Transport from Kubo Formulae,''
  Lect.\ Notes Phys.\  {\bf 871} (2013) 433
  %doi:10.1007/978-3-642-37305-3_17
  [arXiv:1207.5808 [hep-th]].
  %%CITATION = doi:10.1007/978-3-642-37305-3_17;%%
  %55 citations counted in INSPIRE as of 26 janv. 2016

\bibitem{semimetal_effects7}
M. N. Chernodub, A. Cortijo, A. G. Grushin, K. Landsteiner, and M. A. Vozmediano,
{\sl ``A condensed matter realization of the axial magnetic effect''},
Phys. Rev. B {\bf 89}, 081407(R) (2014) [arXiv:1311.0878].


%\cite{Gorbar:2015wya}
\bibitem{Gorbar:2015wya}
  E.~V.~Gorbar, V.~A.~Miransky, I.~A.~Shovkovy and P.~O.~Sukhachov,
  ``Chiral separation and chiral magnetic effects in a slab: The role of boundaries,''
  Phys.\ Rev.\ B {\bf 92} (2015) 24,  245440
  doi:10.1103/PhysRevB.92.245440
  [arXiv:1509.06769 [cond-mat.mes-hall]].
  %%CITATION = doi:10.1103/PhysRevB.92.245440;%%

\bibitem{Miransky:2015ava}
  V.~A.~Miransky and I.~A.~Shovkovy,
  ``Quantum field theory in a magnetic field: From quantum chromodynamics to graphene and Dirac semimetals,''
  Phys.\ Rept.\  {\bf 576} (2015) 1
  %doi:10.1016/j.physrep.2015.02.003
  [arXiv:1503.00732 [hep-ph]].
  %%CITATION = doi:10.1016/j.physrep.2015.02.003;%%
  %46 citations counted in INSPIRE as of 31 janv. 2016

%%%%%%%%%%%%%%%%%%%%%%%%


\bibitem{Valgushev:2015pjn}
  S.~N.~Valgushev, M.~Puhr and P.~V.~Buividovich,
  ``Chiral Magnetic Effect in finite-size samples of parity-breaking Weyl semimetals,''
  arXiv:1512.01405 [cond-mat.str-el].
  %%CITATION = ARXIV:1512.01405;%%
%\cite{Buividovich:2015qba}





\bibitem{Buividovich:2015ara}
  P.~V.~Buividovich, M.~Puhr and S.~N.~Valgushev,
  ``Chiral magnetic conductivity in an interacting lattice model of parity-breaking Weyl semimetal,''
  Phys.\ Rev.\ B {\bf 92} (2015) 20,  205122
  doi:10.1103/PhysRevB.92.205122
  [arXiv:1505.04582 [cond-mat.str-el]].
  %%CITATION = doi:10.1103/PhysRevB.92.205122;%%
  %4 citations counted in INSPIRE as of 31 Jan 2016
%\cite{Kochetkov:2014sxa}


\bibitem{Buividovich:2014dha}
  P.~V.~Buividovich,
  ``Spontaneous chiral symmetry breaking and the Chiral Magnetic Effect for interacting Dirac fermions with chiral imbalance,''
  Phys.\ Rev.\ D {\bf 90} (2014) 125025
  doi:10.1103/PhysRevD.90.125025
  [arXiv:1408.4573 [hep-th]].
  %%CITATION = doi:10.1103/PhysRevD.90.125025;%%
  %4 citations counted in INSPIRE as of 31 Jan 2016
%\cite{Braguta:2014ksa}



\bibitem{Buividovich:2013hza}
  P.~V.~Buividovich,
  ``Anomalous transport with overlap fermions,''
  Nucl.\ Phys.\ A {\bf 925} (2014) 218
  doi:10.1016/j.nuclphysa.2014.02.022
  [arXiv:1312.1843 [hep-lat]].
  %%CITATION = doi:10.1016/j.nuclphysa.2014.02.022;%%
  %23 citations counted in INSPIRE as of 31 Jan 2016


%%%%%%%%%%%%%%%%%%%%%%%%
















%%%%%%%%%%%%%





\bibitem{semimetal_effects6}
S. Parameswaran, T. Grover, D. Abanin, D. Pesin, and A. Vishwanath,
{\sl ``Probing the chiral anomaly with nonlocal transport in Weyl semimetals},
Phys. Rev. X {\bf 4}, 031035 (2014) [arXiv:1306.1234].




\bibitem{semimetal_effects10}
M. Vazifeh and M. Franz,
{\sl ``Electromagnetic response of weyl semimetals''},
Phys. Rev. Lett. {\bf 111}, 027201 (2013) [arXiv:1303.5784].

\bibitem{semimetal_effects11}
Y. Chen, S. Wu, and A. Burkov,
{\sl ``Axion response in Weyl semimetals''},
Phys. Rev. B {\bf 88}, 125105 (2013) [arXiv:1306.5344].

\bibitem{semimetal_effects12}
Y. Chen, D. Bergman, and A. Burkov,
{\sl ``Weyl fermions and the anomalous Hall effect in metallic ferromagnets''},
Phys. Rev. B {\bf 88}, 125110 (2013) [arXiv:1305.0183];
David Vanderbilt, Ivo Souza, and F. D. M. Haldane
Phys. Rev. B {\bf 89}, 117101 (2014) [arXiv:1312.4200].

\bibitem{semimetal_effects13}
S. T. Ramamurthy and T. L. Hughes,
{\sl ``Patterns of electro-magnetic response in topological semi-metals''},
arXiv:1405.7377.









\bibitem{Zyuzin:2012tv}
A.~A.~Zyuzin and A.~A.~Burkov,
  {\sl ``Topological response in Weyl semimetals and the chiral anomaly,''}  Phys.\ Rev.\ B {\bf 86} (2012) 115133  [arXiv:1206.1868 [cond-mat.mes-hall]].  %%CITATION = ARXIV:1206.1868;%%  %37 citations counted in INSPIRE as of 04 Jan 2015

%\bibitem{chiral_torsion3}
%L Sun, S Wan, Chiral viscoelastic response in Weyl semimetals, Europhysics Letters (2014) 108 37007 doi:10.1209/0295-
%5075/108/37007

\bibitem{tewary}
Pallab Goswami, Sumanta Tewari, {\sl Axionic field theory of (3+1)-dimensional Weyl semi-metals,}
Phys. Rev. B 88, 245107 (2013), arXiv:1210.6352

%\cite{Kharzeev:2015znc}
\bibitem{Kharzeev:2015znc}
  D.~E.~Kharzeev, J.~Liao, S.~A.~Voloshin and G.~Wang,
  ``Chiral Magnetic Effect in High-Energy Nuclear Collisions --- A Status Report,''
  arXiv:1511.04050 [hep-ph].
  %%CITATION = ARXIV:1511.04050;%%
  %7 citations counted in INSPIRE as of 31 Jan 2016

\bibitem{Kharzeev:2009mf}
  D.~E.~Kharzeev,
  ``Chern-Simons current and local parity violation in hot QCD matter,''
  Nucl.\ Phys.\ A {\bf 830} (2009) 543C
  doi:10.1016/j.nuclphysa.2009.10.049
  [arXiv:0908.0314 [hep-ph]].
  %%CITATION = doi:10.1016/j.nuclphysa.2009.10.049;%%
  %37 citations counted in INSPIRE as of 31 janv. 2016

\bibitem{Polikarp} P. V. Buividovich, M. N. Chernodub, D. E. Kharzeev, T. Kalaydzhyan,
E. V. Luschevskaya and M. I. Polikarpov, Phys. Rev. Lett. 105, 132001
(2010) [arXiv:1003.2180 [hep-lat]].



 \bibitem{Volovik2003} G.E. Volovik, {\it The Universe in a Helium
Droplet}, Clarendon Press,  Oxford (2003).

%\cite{Volovik:2011kg}
\bibitem{Volovik:2011kg}
  G.~E.~Volovik,
  ``Topology of quantum vacuum,''
  Lecture Notes in Physics {\bf 870} (2013) 343
  [arXiv:1111.4627 [hep-ph]].
  %%CITATION = ARXIV:1111.4627;%%
  %29 citations counted in INSPIRE as of 10 Feb 2016

\bibitem{VZ2012}
M.A.Zubkov, G.E.Volovik, Momentum space topological invariants for the $4D$
relativistic vacua with mass gap, Nuclear Physics B (2012)
doi:10.1016/j.nuclphysb.2012.03.002, ArXiv:1201.4185


\bibitem{Z2012} M.A.Zubkov, Generalized unparticles, zeros of the Green function, and momentum space topology of the lattice model with overlap fermions,  Phys. Rev. D 86, 034505 (2012)

\bibitem{Wigner}  E.P. Wigner, "On the quantum correction for thermodynamic equilibrium", Phys. Rev. 40 (June 1932) 749–759. doi:10.1103/PhysRev.40.749

\bibitem{star} C. Zachos, D. Fairlie, and T. Curtright, "Quantum Mechanics in Phase Space" ( World Scientific, Singapore, 2005) ISBN 978-981-238-384-6

\bibitem{star2} 	
F. Bayen, M. Flato, C. Fronsdal, A. Lichnerowicz, D. Sternheimer, "Deformation Theory and Quantization. 2. Physical Applications",
 Annals Phys. 111 (1978) 111

 \bibitem{ke}
 D.~T.~Son and N.~Yamamoto,
  ``Kinetic theory with Berry curvature from quantum field theories,''
  Phys.\ Rev.\ D {\bf 87} (2013) no.8,  085016
  doi:10.1103/PhysRevD.87.085016
  [arXiv:1210.8158 [hep-th]].
  %%CITATION = doi:10.1103/PhysRevD.87.085016;%%
  %77 citations counted in INSPIRE as of 25 Mar 2016

\bibitem{ke2}
  J.~W.~Chen, S.~Pu, Q.~Wang and X.~N.~Wang,
  ``Berry Curvature and Four-Dimensional Monopoles in the Relativistic Chiral Kinetic Equation,''
  Phys.\ Rev.\ Lett.\  {\bf 110}, no. 26, 262301 (2013)
  doi:10.1103/PhysRevLett.110.262301
  [arXiv:1210.8312 [hep-th]].
  %%CITATION = doi:10.1103/PhysRevLett.110.262301;%%
  %68 citations counted in INSPIRE as of 25 Mar 2016

\bibitem{ke3}
  J.~H.~Gao, Z.~T.~Liang, S.~Pu, Q.~Wang and X.~N.~Wang,
  ``Chiral Anomaly and Local Polarization Effect from Quantum Kinetic Approach,''
  Phys.\ Rev.\ Lett.\  {\bf 109} (2012) 232301
  doi:10.1103/PhysRevLett.109.232301
  [arXiv:1203.0725 [hep-ph]].
  %%CITATION = doi:10.1103/PhysRevLett.109.232301;%%
  %62 citations counted in INSPIRE as of 25 Mar 2016


\bibitem{anomaly_U}
Shifman M. A. "Anomalies in Gauge Theories", Phys. Rep. 209 (1991), 341



\bibitem{Volovik1988} G.E.Volovik, "Analogue of quantum Hall effect in superfluid $^3$He  film" JETP 67, 1804 - 1811

\bibitem{nogo}
 M. M. Vazifeh and M. Franz, Phys. Rev. Lett. 111,
027201 (2013)

\bibitem{nogo2} N. Yamamoto, Phys. Rev. D 92, 085011 (2015).

\bibitem{Weyl}
Robert G Littlejohn, "The semiclassical evolution of wave packets",
Physics Reports Volume 138, Issues 4–5, May 1986, Pages 193-291

\bibitem{berezin} Berezin, F.A. and M.A. Shubin, 1972, in: Colloquia Mathematica Societatis Janos Bolyai (North-Holland, Amsterdam) p. 21.


\bibitem{Katsbook} M.I. Katsnelson, Graphene: Carbon in Two
Dimensions, Cambridge Univ. Press, Cambridge, 2012.

\bibitem{Hou:2011ze}
  D.~Hou, H.~Liu and H.~c.~Ren,
  ``Some Field Theoretic Issues Regarding the Chiral Magnetic Effect,''
  JHEP {\bf 1105} (2011) 046
  doi:10.1007/JHEP05(2011)046
  [arXiv:1103.2035 [hep-ph]].
  %%CITATION = doi:10.1007/JHEP05(2011)046;%%
  %28 citations counted in INSPIRE as of 02 May 2016

\bibitem{Rebhan}
A. Rebhan, A. Schmitt and S. A. Stricker, “Anomalies and the chiral magnetic
effect in the Sakai-Sugimoto model”, JHEP 1001, 026 (2010); arXiv:
0909.4782[hep-th].

\bibitem{Rubakov}
 V.~A.~Rubakov,
  ``On chiral magnetic effect and holography,''
  arXiv:1005.1888 [hep-ph].
  %%CITATION = ARXIV:1005.1888;%%
  %66 citations counted in INSPIRE as of 02 May 2016

\bibitem{Gorsky}
A. Gorsky, P. N. Kopnin and A. V. Zayakin, On the Chiral Magnetic Effect in Soft-Wall AdS/QCD,
Phys. Rev. D83 (2011) 014023 [arXiv:1003.2293]


\end{thebibliography}
\end{document}